\documentclass[sn-nature]{sn-jnl}

\usepackage{graphicx}
\graphicspath{{figures/}}

\usepackage{multirow}
\usepackage{amsmath,amssymb,amsfonts,amsthm}
\usepackage{mathrsfs}
\usepackage[title]{appendix}
\usepackage{xcolor}
\usepackage{textcomp}
\usepackage{manyfoot}
\usepackage{placeins}
\usepackage{booktabs}
\usepackage{listings}

\newcommand{\apjl}{Astrophys. J. Lett.}   
\newcommand{\apjs}{Astrophys. J. Suppl. Ser.}   
\newcommand{\ao}{Appl. Opt.}   
\newcommand{\aap}{Astron. Astrophys.}   

\usepackage{siunitx}[=v2]
\DeclareSIUnit{\year}{yr}
\DeclareSIUnit{\pixel}{pix}
\DeclareSIUnit{\arcmin}{arcmin}
\DeclareSIUnit{\parsec}{pc}
\DeclareSIUnit{\gauss}{G}
\DeclareSIUnit{\erg}{erg}
\DeclareSIUnit{\mas}{mas} 
\DeclareSIUnit{\arcsec}{arcsec}
\DeclareSIUnit{\jansky}{Jy}
\DeclareSIUnit{\FRBs}{FRBs}
\DeclareSIUnit{\FRB}{FRB}
\DeclareSIUnit{\sky}{sky}
\DeclareSIUnit{\day}{day}
\DeclareSIUnit{\month}{month}
\DeclareSIUnit{\ethernet}{E}
\DeclareSIUnit{\TECu}{TECu}
\DeclareSIUnit{\electron}{electron}
\DeclareSIUnit{\MSPS}{MSPS}

\usepackage[super]{nth}
\usepackage{upgreek}
\usepackage{xspace}
\usepackage{tabularx}
\usepackage{commath}

\newcommand{\dmfrb}{\SI{500.147 +- 0.004}{\parsec\per\centi\m\cubed}\xspace}
\newcommand{\redshift}{\num{0.1772 +- 0.0001}\xspace}
\newcommand{\dmmwne}{\SI{40 +- 8}{\parsec\per\centi\m\cubed}\xspace}
\newcommand{\dmmwhalo}{\SI{30 +- 20}{\parsec\per\centi\m\cubed}\xspace}
\newcommand{\dmcosmic}{\SI{172 +- 90}{\parsec\per\centi\m\cubed}\xspace}
\newcommand{\inclination}{\SI{7 +- 3}{\degree}}
\newcommand{\geometryfactor}{\num{8 +- 3}}
\newcommand{\dmhostbudget}{\SI{257 +- 93}{\parsec\per\centi\m\cubed}\xspace}
\newcommand{\dmhostbudgetrest}{\SI{302 +- 109}{\parsec\per\centi\m\cubed}\xspace}
\newcommand{\dmhostdiskrest}{\SI{193 +- 82}{\parsec\per\centi\m\cubed}\xspace}
\newcommand{\dmhosthalorest}{\SI{48 +- 23}{\parsec\per\centi\m\cubed}\xspace}
\newcommand{\dmhostestimaterest}{\SI{264 +- 97}{\parsec\per\centi\m\cubed}\xspace}
\newcommand{\dmhostestimate}{\SI{224 +- 82}{\parsec\per\centi\m\cubed}\xspace}
\newcommand{\dmhostestimatebehind}{\SI{448 +- 164}{\parsec\per\centi\m\cubed}\xspace}
\newcommand{\dmhostdisksubtractrest}{\SI{254 +- 111}{\parsec\per\centi\m\cubed}\xspace}
\newcommand{\dmmwhaloavg}{\SI{43 +- 20}{\parsec\per\centi\m\cubed}\xspace}
\newcommand{\dmmwperp}{\SI{24 +- 3}{\parsec\per\centi\m\cubed}\xspace}
\newcommand{\massmw}{\num{6.1 +- 1.1}}
\newcommand{\masshost}{\num{8.5 +- 0.8}}
\newcommand{\massratio}{\num{1.4 +- 0.3}}
\newcommand{\massscalingfactor}{\num{1.12 +- 0.08}}
\newcommand{\tauscmodel}{\SI{1.0 +- 0.5}{\micro\second}\xspace}
\newcommand{\rmexcess}{\SI[parse-numbers=false]{(+198 \pm 3)}{\radian\per\meter\squared}\xspace}

\newcommand{\Msun}{\ensuremath{~\text{M}_\odot}}
\newcommand{\tentoten}{\ensuremath{\times 10^{10}}}
\newcommand{\rahms}{$0\text{h}41\text{m}05.774\text{s}\pm0.0192\text{s}$\xspace}
\newcommand{\decdms}{$21\text{d}13\text{m}34.573\text{s}\pm1.08\text{s}$\xspace}
\newcommand{\radecimal}{\SI{10.274058 +- 0.00008}{\degree}} 
\newcommand{\decdecimal}{\SI{21.226270 +- 0.0003}{\degree}} 
\newcommand{\dustcorr}{1.97}

\usepackage{hyperref}
\hypersetup{
    pdffitwindow=true,
    pdfstartview={FitH},
    pdftitle={A fast radio burst localized at detection to an edge-on galaxy using very-long-baseline interferometry},
    pdfauthor={Tomas Cassanelli, Calvin Leung, Pranav Sanghavi},
    pdfcreator={Tomas Cassanelli, Calvin Leung, Pranav Sanghavi},
    pdfproducer={Tomas Cassanelli, Calvin Leung, Pranav Sanghavi},
    pdfkeywords={
        Astronomy, Astrophysics, VLBI, Interferometry, Telescope, Radio
        },
    pdfnewwindow=true,
    colorlinks=true,
    linkcolor=blue,
    citecolor=magenta,
    urlcolor=magenta
    }

\begin{document}

\title[A fast radio burst localized at detection to an edge-on galaxy using very-long-baseline interferometry]{A fast radio burst localized at detection to an edge-on galaxy using very-long-baseline interferometry}








\author[1,2,3]{\fnm{Tomas} \sur{Cassanelli}}\email{tcassanelli@ing.uchile.cl}
\equalcont{These authors contributed equally to this work.}

\author*[4,5,6]{\fnm{Calvin} \sur{Leung}}\email{calvin\_leung@berkeley.edu}
\equalcont{These authors contributed equally to this work.}

\author[7,8,9]{\fnm{Pranav} \sur{Sanghavi}}\email{pranav.sanghavi@yale.edu}
\equalcont{These authors contributed equally to this work.}

\author[1,2]{\fnm{Juan} \sur{Mena-Parra}}

\author[4,10]{\fnm{Savannah} \sur{Cary}}

\author[11,12]{\fnm{Ryan} \sur{Mckinven}}

\author[11,12]{\fnm{Mohit} \sur{Bhardwaj}}

\author[4,5]{\fnm{Kiyoshi W.} \sur{Masui}}

\author[4,5]{\fnm{Daniele} \sur{Michilli}}

\author[7,8]{\fnm{Kevin} \sur{Bandura}}

\author[13]{\fnm{Shami} \sur{Chatterjee}}

\author[14]{\fnm{Jeffrey B.} \sur{Peterson}}

\author[15,16,17]{\fnm{Jane} \sur{Kaczmarek}}

\author[18]{\fnm{Mubdi} \sur{Rahman}}

\author[4,5]{\fnm{Kaitlyn} \sur{Shin}}

\author[1,2]{\fnm{Keith} \sur{Vanderlinde}}

\author[11,12]{\fnm{Sabrina} \sur{Berger}}

\author[11,12]{\fnm{Charanjot} \sur{Brar}}

\author[11,12]{\fnm{P.~J.} \sur{Boyle}}

\author[19]{\fnm{Daniela} \sur{Breitman}}

\author[20]{\fnm{Pragya} \sur{Chawla}}

\author[11,12]{\fnm{Alice P.} \sur{Curtin}}

\author[11,12]{\fnm{Matt} \sur{Dobbs}}

\author[21]{\fnm{Fengqiu Adam} \sur{Dong}}

\author[8,22]{\fnm{Emmanuel} \sur{Fonseca}}

\author[1,2,23]{\fnm{B.~M.} \sur{Gaensler}}

\author[1,2]{\fnm{Adaeze} \sur{Ibik}}

\author[11,12]{\fnm{Victoria~M.} \sur{Kaspi}}

\author[7,8]{\fnm{Khairy} \sur{Kholoud}}

\author[11,12]{\fnm{Adam E.} \sur{Lanman}}

\author[1, 11]{\fnm{Mattias} \sur{Lazda}}

\author[24,25]{\fnm{Hsiu-Hsien} \sur{Lin}}

\author[26]{\fnm{Jing} \sur{Luo}}

\author[21]{\fnm{Bradley W.} \sur{Meyers}}

\author[21]{\fnm{Nikola} \sur{Milutinovic}}

\author[2]{\fnm{Cherry} \sur{Ng}}

\author[1,2]{\fnm{Gavin} \sur{Noble}}

\author[11,12,27,28,29]{\fnm{Aaron~B.} \sur{Pearlman}}

\author[2,24,25,30,31]{\fnm{Ue-Li} \sur{Pen}}

\author[11,12,20]{\fnm{Emily} \sur{Petroff}}

\author[2]{\fnm{Ziggy} \sur{Pleunis}}

\author[32,33]{\fnm{Brendan} \sur{Quine}}

\author[11,12]{\fnm{Masoud} \sur{Rafiei-Ravandi}}

\author[2]{\fnm{Andre} \sur{Renard}}

\author[11,12]{\fnm{Ketan R.} \sur{Sand}}

\author[4]{\fnm{Eve} \sur{Schoen}}

\author[2]{\fnm{Paul} \sur{Scholz}}

\author[31]{\fnm{Kendrick M.} \sur{Smith}}

\author[21]{\fnm{Ingrid} \sur{Stairs}}

\author[34,35]{\fnm{Shriharsh P.} \sur{Tendulkar}}


\affil[1]{\orgname{David A.~Dunlap Department of Astronomy \& Astrophysics, University of Toronto}, \orgaddress{\street{50 St.~George Street}, \city{Toronto}, \postcode{M5S 3H4}, \state{ON}, \country{Canada}}}

\affil[2]{\orgname{Dunlap Institute for Astronomy \& Astrophysics, University of Toronto}, \orgaddress{\street{50 St.~George Street}, \city{Toronto}, \postcode{M5S 3H4, \state{ON}, \country{Canada}}}}

\affil[3]{\orgname{Department of Electrical Engineering, Universidad de Chile}, \orgaddress{\street{Av. Tupper 2007}, \city{Santiago}, \postcode{8370451}, \state{}, \country{Chile}}}

\affil*[4]{\orgname{MIT Kavli Institute for Astrophysics and Space Research, Massachusetts Institute of Technology}, \orgaddress{\street{77 Massachusetts Ave}, \city{Cambridge}, \postcode{02139}, \state{MA}, \country{USA}}}

\affil*[5]{\orgname{Department of Physics, Massachusetts Institute of Technology}, \orgaddress{\street{77 Massachusetts Ave}, \city{Cambridge}, \postcode{02139}, \state{MA}, \country{USA}}}

\affil*[6]{\orgname{NHFP Einstein Fellow}}

\affil[7]{\orgname{Lane Department of Computer Science and Electrical Engineering, West Virginia University}, \orgaddress{\street{1220 Evansdale Drive, PO Box 6109}, \city{Morgantown}, \postcode{26506}, \state{WV}, \country{USA}}}

\affil[8]{\orgname{Center for Gravitational Waves and Cosmology, West Virginia University, Chestnut Ridge Research Building}, \orgaddress{\city{Morgantown}, \postcode{26505}, \state{WV}, \country{USA}}}

\affil[9]{\orgname{Department of Physics, Yale University}, \orgaddress{\city{New Haven}, \postcode{06520}, \state{CT}, \country{USA}}}

\affil[10]{\orgname{Department of Astronomy, Wellesley College}, \orgaddress{\street{106 Central Street}, \city{Wellesley}, \postcode{02481}, \state{MA}, \country{USA}}}

\affil[11]{\orgname{Department of Physics, McGill University}, \orgaddress{\street{3600 rue University}, \city{Montr\'eal}, \postcode{H3A 2T8}, \state{QC}, \country{Canada}}}

\affil[12]{\orgname{Trottier Space Institute, McGill University}, \orgaddress{\street{3550 rue University}, \city{Montr\'eal}, \postcode{H3A 2A7}, \state{QC}, \country{Canada}}}

\affil[13]{\orgname{Cornell Center for Astrophysics and Planetary Science}, \orgaddress{\city{Ithaca}, \postcode{14853}, \state{NY}, \country{USA}}}

\affil[14]{\orgname{McWilliams Center for Cosmology, Department of Physics, Carnegie Mellon University}, \orgaddress{\street{5000 Forbes Ave}, \city{Pittsburgh}, \postcode{15213}, \state{PA}, \country{USA}}}

\affil[15]{\orgname{Dominion Radio Astrophysical Observatory, Herzberg Research Centre for Astronomy and Astrophysics, National Research Council Canada, PO Box 248}, \orgaddress{\city{Penticton}, \postcode{V2A 6J9}, \state{BC}, \country{Canada}}}

\affil[16]{\orgname{Department of Computer Science, Math, Physics, \& Statistics, University of British Columbia}, \orgaddress{\city{Kelowna}, \postcode{V1V 1V7}, \state{BC}, \country{Canada}}}

\affil[17]{\orgname{CSIRO Space \& Astronomy, Parkes Observatory, P.O. Box 276}, \orgaddress{\city{Parkes}, \postcode{2870}, \state{NSW}, \country{Australia}}}

\affil[18]{\orgname{Sidrat Research}, \orgaddress{\street{124 Merton Street, Suite 507}, \city{Toronto}, \postcode{M4S 2Z2}, \state{ON}, \country{Canada}}}

\affil[19]{\orgname{Scuola Normale Superiore}, \orgaddress{\city{Pisa}, \postcode{56126}, \state{PI}, \country{Italy}}}

\affil[20]{\orgname{Anton Pannekoek Institute for Astronomy, University of Amsterdam}, \orgaddress{\street{Science Park 904}, \city{Amsterdam}, \postcode{1098 XH}, \country{The Netherlands}}}

\affil[21]{\orgname{Department of Physics and Astronomy, University of British Columbia}, \orgaddress{\street{6224 Agricultural Road}, \city{Vancouver}, \postcode{V6T 1Z1}, \state{BC}, \country{Canada}}}

\affil[22]{\orgname{Department of Physics and Astronomy, West Virginia University, P.O. Box 6315}, \orgaddress{\city{Morgantown}, \postcode{26506}, \state{WV}, \country{USA}}}

\affil[23]{\orgname{Department of Astronomy and Astrophysics, University of California Santa Cruz}, \orgaddress{\street{1156 High Street}, \city{Santa Cruz}, \postcode{95064}, \state{CA}, \country{USA}}}

\affil[24]{\orgname{Institute of Astronomy and Astrophysics, Academia Sinica}, \orgaddress{\street{Astronomy-Mathematics Building, No. 1, Sec. 4, Roosevelt Road}, \city{Taipei}, \postcode{10617}, \country{Taiwan}}}

\affil[25]{\orgname{Canadian Institute for Theoretical Astrophysics}, \orgaddress{\street{60 St.~George Street}, \city{Toronto}, \postcode{M5S 3H8}, \state{ON}, \country{Canada}}}

\affil[26]{\orgname{Deceased}}

\affil[27]{\orgname{Banting Fellow}}

\affil[28]{\orgname{McGill Space Institute Fellow}}

\affil[29]{\orgname{FRQNT Postdoctoral Fellow}}

\affil[30]{\orgname{Canadian Institute for Advanced Research}, \orgaddress{\street{661 University Ave}, \city{Toronto}, \postcode{M5G 1M1}, \state{ON}, \country{Canada}}}

\affil[31]{\orgname{Perimeter Institute for Theoretical Physics}, \orgaddress{\street{31 Caroline Street N}, \city{Waterloo}, \postcode{N25 2YL}, \state{ON}, \country{Canada}}}

\affil[32]{\orgname{Thoth Technology Inc.}, \orgaddress{\street{33387 Highway 17}, \city{Deep River}, \postcode{K0J 1P0}, \state{ON}, \country{Canada}}}

\affil[33]{\orgname{Department of Physics and Astronomy, York University}, \orgaddress{\street{4700 Keele Street}, \city{Toronto}, \postcode{M3J 1P3}, \state{ON}, \country{Canada}}}

\affil[34]{\orgname{Department of Astronomy and Astrophysics, Tata Institute of Fundamental Research}, \orgaddress{\city{Mumbai}, \postcode{400005}, \country{India}}}

\affil[35]{\orgname{National Centre for Radio Astrophysics}, \orgaddress{\street{Post Bag 3, Ganeshkhind}, \city{Pune}, \postcode{411007}, \country{India}}}


\newcommand{\allacks}{
A.B.P. is a Banting Fellow, a McGill Space Institute~(MSI) Fellow, and a Fonds de Recherche du Quebec -- Nature et Technologies~(FRQNT) postdoctoral fellow.
A.P.C. is a Vanier Canada Graduate Scholar.
B.M.G. acknowledges the support of the Natural Sciences and Engineering Research Council of Canada (NSERC) through grant RGPIN-2015-05948, and of the Canada Research Chairs program.
C.L. was supported by the U.S. Department of Defense (DoD) through the National Defense Science \& Engineering Graduate Fellowship (NDSEG) Program.
E.P. acknowledges funding from an NWO Veni Fellowship.
F.A.D is funded by the UBC Four Year Doctoral Fellowship.
FRB research at UBC is funded by an NSERC Discovery Grant and by the Canadian Institute for Advance Research. The CHIME baseband system was funded in part by a CFI JELF award to IHS. 
FRB research at WVU is supported by an NSF grant (2006548, 2018490)
J.B.P. is support by the NSF MRI grant (2018490)
J.M.P is a Kavli Fellow.
K.S. is supported by the NSF Graduate Research Fellowship Program.
K.W.M. is supported by an NSF Grant (2008031).
M.B. is supported by an FRQNT Doctoral Research Award.
M.D. is supported by a Canada Research Chair, Killam Fellowship, NSERC Discovery Grant, CIFAR, and by the FRQNT Centre de Recherche en Astrophysique du Qu\'ebec (CRAQ)
P.S. is a Dunlap Fellow. 
S. Cary would like to thank Prof. Kim McLeod from Wellesley College for her supervision and feedback, which was essential for the host galaxy analysis.
S.C. is a member of the NANOGrav Physics Frontiers Center, which is supported by NSF award PHY-1430284.
U.P. is supported by the Natural Sciences and Engineering Research Council of Canada (NSERC), [funding reference number RGPIN-2019-067, CRD 523638-201, 555585-20], Ontario Research Fund—research Excellence Program (ORF-RE), Canadian Institute for Advanced Research (CIFAR), Simons Foundation, Thoth Technology Inc, and Alexander von Humboldt Foundation. This research is supported by the MOST grant 110-2112-M-001-071-MY3 from the Ministry of Science and Technology of Taiwan. 
V.M.K. holds the Lorne Trottier Chair in Astrophysics \& Cosmology, a Distinguished James McGill Professorship, and receives support from an NSERC Discovery grant (RGPIN 228738-13), from an R. Howard Webster Foundation Fellowship from CIFAR, and from the FRQNT CRAQ.
Z.P. is a Dunlap Fellow.
}


\abstract{Fast radio bursts (FRBs) are millisecond-duration radio transients whose origins remain unknown. Since the vast majority of bursts are one-off events, it is necessary to pinpoint FRBs precisely within their host galaxies at the time of detection. Here, we use two purpose-built outrigger telescopes to localize FRB 20210603A at the time of its detection by the Canadian Hydrogen Intensity Mapping Experiment (CHIME). Our VLBI stations localize the burst to a $\ang{;;.2}\times\ang{;;2}$ final ellipse in the disk of its host galaxy SDSS J004105.82+211331.9. Spatially-resolved spectroscopic followup reveals recent star formation (H$\alpha$ emission) on kiloparsec scales near the burst position. The excess DM is consistent with expectations from the nearly edge-on disk of the host galaxy, demonstrating the utility of FRBs as probes of the interstellar medium in distant galaxies. The excess DM, rotation measure, and scattering are consistent with expectations for a pulse traveling from deep within its host galactic plane, strengthening the link between the local environment of FRB 20210603A and the disk of its host galaxy. Finally, this technique demonstrates a way to overcome the trade-off between angular resolution and field of view in FRB instrumentation, paving the way towards plentiful and precise FRB localizations.}




\maketitle

\begin{figure}
    \centering
    \includegraphics{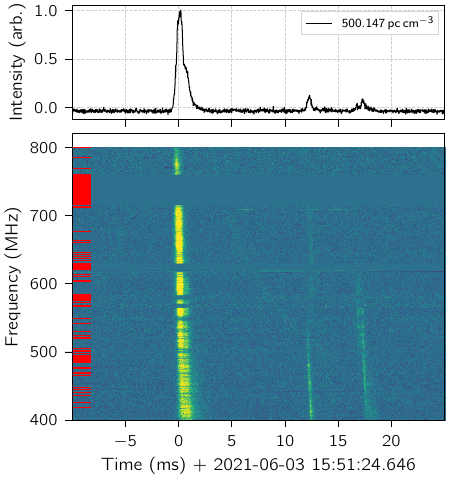}
    \caption{\textbf{The Stokes-$I$ dynamic spectrum of FRB 20210603A.} We detect the single pulse in autocorrelation at CHIME/FRB with a signal-to-noise ratio exceeding \num{100}. The data are shown at a time resolution of \SI{25.6}{\micro\s} with pixel colours scaled to their \numrange{1}{99} percentile values. To remove dispersion, we use a DM derived by lining up three closely-overlapping sub-burst components within the main pulse using \texttt{fitburst} \cite{2021ApJS..257...59C,2024ApJS..271...49F}. In addition to the main burst, fainter emission components occurring \SI{\sim12}{\milli\s} and \SI{\sim18}{\milli\s} afterwards are visible in CHIME/FRB baseband data, but are neglected for VLBI localization. The faint dispersed sweeps left and right of the main pulse are known instrumental artifacts from spectral leakage.
    The red streaks to the left highlight the frequency channels that are masked out due to RFI. Most RFI come from cellular communication and television transmission bands between \SIrange{700}{750}{\mega\hertz} and \SIrange{600}{650}{\mega\hertz}, respectively.}
    \label{fig:frb_i_stokes}
\end{figure}

\begin{figure}
    \centering
    \includegraphics{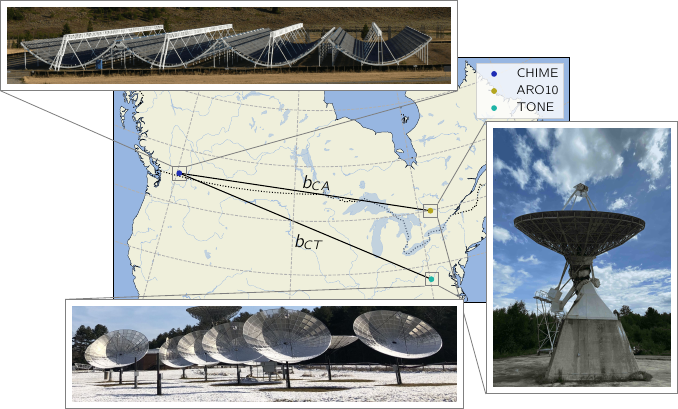}
    \caption{\textbf{Map of baselines formed between CHIME and ARO10 (CA) and TONE (CT).} The baselines span from Penticton, BC to Algonquin, ON, and Green Bank, WV with lengths $b_\mathrm{CA}=\SI{3074}{\kilo\m}$ and $b_\mathrm{CT}=\SI{3332}{\kilo\m}$. For our localization analysis, we omit the \SI{848}{\kilo\m} baseline between ARO10 and TONE because the FRB was not sufficiently bright to be detected on that baseline. Properties of each station are listed in Extended Data Table \ref{tab:telescope_properties}.}
    \label{fig:vlbi_projection}
\end{figure}

\begin{figure}
    \centering
    \includegraphics{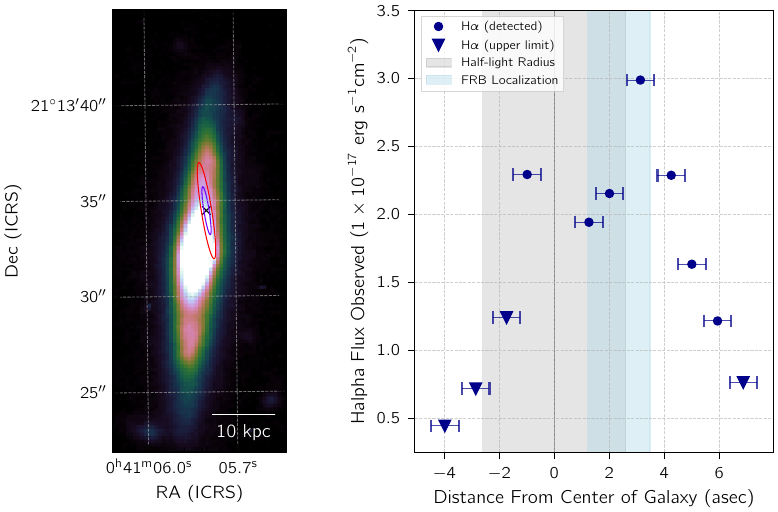}
    \caption{\textbf{VLBI Localization of FRB 20210603A}. Left: The $1\sigma$ and $2\sigma$ localization contours, defined by an empirical estimate of our localization errors using Crab measurements, are overlaid on a CFHT MegaCAM $gri$-band image of its host galaxy SDSS J004105.82+211331.9. The nearly edge-on geometry of the host galaxy is apparent. We allow the pixel colours to saturate within $\approx 1$ half-light radius, to accentuate the faint structure on the outskirts of the galaxy. The localization and burst properties point towards a progenitor living deep in the ionized disk of the galaxy.
    Right: H$\alpha$ flux observed at varying distances from the galactic center along the major axis of the galaxy, calculated from the spectra in Extended Data Fig.~\ref{fig:galaxy_spectra}. Positive/negative coordinates refer to H$\alpha$ fluxes northward/southward of the galactic center respectively. Blue circles and upside-down triangles represent detections and $2 \sigma$ upper limits on the local H$\alpha$ flux, with flux uncertainties estimated using the detrended spectrum (SD; $N = 3199$).
    Horizontal bars denote the size of the spectral aperture ($\SI{1}{\arcsec}$). The half light radius of the galaxy is indicated by a gray shaded area.
    }
    \label{fig:galaxy_image}
\end{figure}

FRB 20210603A (Fig.~\ref{fig:frb_i_stokes}) was first detected by the FRB search backend of the Canadian Hydrogen Intensity Mapping Experiment (CHIME)~\cite{2018ApJ...863...48C}, which searches for dispersed single pulses within search beams tiling the \SI{200}{\deg\squared} primary beam of the telescope. The high signal-to-noise ratio of the burst ($>\num{100}$) triggered the recording of voltage data at CHIME~\cite{2022ApJS..261...29C} and two small telescopes shadowing a portion of the CHIME FoV: a 10-m single dish at Algonquin Radio Observatory (hereafter, ARO10) \cite{2022AJ....163...65C}, and TONE, a compact array of eight, 6-m dishes at Green Bank Observatory (GBO)\cite{tonesystem}. Voltage data dumps from all three stations (Fig.~\ref{fig:vlbi_projection}) as well as daily calibration dumps on the Crab pulsar enabled the FRB to be pinpointed to the host galaxy SDSS J004105.82+211331.9 (Fig.~\ref{fig:galaxy_image}).

The three stations in our ad-hoc array are fixed and share a common field of view. CHIME is a compact interferometer with 1024 antennas whose field of view consists of a $\SI{\sim 110}{\degree} \times \SI{\sim 2}{\degree}$ strip aligned along the local meridian~\cite{2018ApJ...863...48C}. Like CHIME, ARO10 and TONE drift-scan the sky. They are manually pointed such that their common field of view overlaps the CHIME/FRB search beams at a declination of $\SI{\sim +22}{\degree}$ such that the Crab pulsar can be used as a VLBI calibrator. When a search beam within the common field of view detects a sufficiently-bright single pulse, low-latency alerts trigger dumps of data across the VLBI network (see Methods: Instrumentation and Observations). This observing mode is a technology demonstration for CHIME/FRB Outriggers, which will expand the strategy to the full field of view of CHIME using more sensitive outrigger telescopes. Owing to the low sensitivity of ARO10 and TONE, our ad-hoc VLBI array only has two useful baselines. Both are largely east-west baselines; nevertheless they together provide sufficient (arcsecond-scale) resolution in the north-south direction. 
The short internal baselines within CHIME and TONE do not contain much astrometric information. Since they are much shorter than the long inter-station baselines, we form voltage beams at CHIME and TONE towards the single pulses of interest. Cable delays for each antenna within CHIME are calculated using the calibration solutions from the CHIME 21-cm backend~\cite{2021ApJ...910..147M}; for TONE we use scheduled voltage dumps from the daily transit of Tau A (the Crab nebula) to measure cable delays~\cite{tonesystem}. Applying these delays allows station beams to be formed towards the best-fit position obtained using CHIME only, which we obtain using the baseband localization pipeline (see Methods: Local calibration and beamforming). We refer to this arcminute-precision position as $\widehat{\mathbf n}_0$.
After forming station beams, our custom-written VLBI correlator~\cite{2024arXiv240305631L} takes the voltage data from beamformed CHIME, beamformed TONE, and ARO10. Within the correlator, geometric delays and Doppler corrections from the Consensus model \cite{eubanks_proceedings_1991} are applied to the voltage data in each of the \num{1024} frequency channels. We omit ionospheric and clock corrections from the delay model, and calibrate these effects out at the level of visibilities. Our correlator then applies coherent dedispersion to the Doppler-corrected voltage data from each station. This reduces the effect of intra-channel smearing and narrows the pulse in time by a factor of a few, which significantly increases the sensitivity of our offline system with respect to the FRB search engine. VLBI correlation allows the FRB signal to be pulled out of the noise at the less sensitive stations, where the FRB is undetectable in auto-correlation. After coherent dedispersion and gating, our long-baseline visibilities are generated and written to disk. Despite being undetectable at the outrigger stations in autocorrelation, the FRB is strongly detected with a $\text{S/N}\sim 35$ in our visibilities on both the CHIME-ARO10 and CHIME-TONE baselines.

After the burst is detected in cross-correlation, ionosphere and clock corrections remain to be applied. Typically, these calibration solutions are straightforward to determine using VLBI observations of continuum sources with precisely-known positions in the same observing session. However, with our ad-hoc array, such observations are difficult due to the unknown availability of VLBI calibrators at \SI{600}{\mega\hertz}, the fixed pointings and low sensitivity of ARO10 and TONE, and the limited internet connectivity of the ARO10 station. Our calibration strategy instead relies on observing bright Crab giant pulses (see Extended Data Fig.~\ref{fig:i_stokes_events}), once per day, resulting in calibration measurements which are much sparser than typically achievable with a mature, steerable VLBI array. Nevertheless, with each baseline individually, we conduct monitoring campaigns of the Crab, where we observed and delay-calibrated 10 Crab GP datasets on the CHIME-ARO10 baseline and 11 on the CHIME-TONE baseline to empirically estimate our $1\sigma$ localization uncertainties (see Extended Data Fig.~\ref{fig:empirical_error}). Since the Crab emits giant pulses unpredictably, we observe them in our system with a range of fluences, spectral properties, and sky locations: in both the CHIME-ARO10 and the CHIME-TONE monitoring campaigns, the pulses spanned a range of $\approx 1.1$ degrees in hour angle. Because our drift-scan telescopes do not track any particular RA, the sky rotation and pulse-to-pulse variability mimics the observation of astrophysical sources with distinct source properties at distinct RAs. The delay uncertainties correspond to a systematic uncertainty ellipse of $\SI{0.2}{\arcsec} \times \SI{2}{\arcsec}$ in the east-west and north-south directions respectively (see Methods: VLBI Calibration and Empirical Localization Error Budget).

In the science run, both the CHIME-ARO10 and the CHIME-TONE baselines operated simultaneously. During this science run we observed FRB 20210603A, and several Crab GP before and after its detection, which we refer to as C1-C4. These GPs allow us to derive a set of phase, delay, and delay-rate calibration solutions, which we used to localize the FRB (Methods: FRB Localization). However, before performing the localization, we validate the calibration solutions by using them to localize a Crab GP (referred to as C3), which we detected one day after the FRB and omitted from our calibration solutions, making it an independent check of our calibration (see Extended Data Figs.~\ref{fig:visibilities_crab},~ \ref{fig:mcmc_corner_c3}, and~\ref{fig:c3_loc}). The discrepancy between the Crab's true position and our Crab localization falls well within the systematic uncertainty ellipse from the monitoring campaigns. Finally, we apply the exact same calibration solutions to localize the FRB (see Extended Data Figs.~\ref{fig:visibilities_frb} and~\ref{fig:mcmc_corner_frb}). The target-calibrator separation is \num{1.5} degrees in hour angle, \num{0.8} degrees in declination, and 4 hours in time. The derived coordinates of FRB 20210603A in the International Celestial Reference Frame (ICRF) are $(\upalpha) =$ \rahms and $(\updelta) =$ +\decdms (Table~\ref{table:frbparams}). These coordinates coincide with SDSS J004105.82+211331.9 \cite{SDSS12}, a disk galaxy with a nearly edge-on orientation (see Figure~\ref{fig:galaxy_image}).

We observed SDSS J004105.82+211331.9 with the Canada-France Hawaii-Telescope (CFHT) MegaCam on 2021 September \nth{10} using the wideband \textit{gri} filter~\cite{1998SPIE.3355..614B}. Figure~\ref{fig:galaxy_image} shows the location of the FRB within the host galaxy. In contrast to other FRB host galaxies that have been robustly identified so far, this galaxy is viewed nearly edge-on; it has an inclination of $\SI{83 +- 3}{\degree}$ (InclinationZoo, \cite{2020ApJ...902..145K}). We determine the \textit{r}-band half-light radius and Galactic extinction-corrected apparent magnitude to be \SI{8.2 +- .9}{\kilo\parsec} and \num{17.90 \pm 0.01}, respectively, using photometric data provided by the Sloan Digital Sky Survey (SDSS~\cite{SDSS12}), see Methods: Host Galaxy Analysis. 

Additionally, we acquired spatially-resolved spectra with the Gemini Multi-Object Spectrograph (GMOS \cite{2004PASP..116..425H}) on 2021 August \nth{1} with the combination of a R400 grating and a GG455 low-pass filter configured with a \SI{1.5}{\arcsec} slit, covering the wavelength range from \SIrange{4650}{8900}{\angstrom}. The slit was co-aligned with the major axis of the galaxy to provide one-dimensional spatial information (Extended Data Fig.~\ref{fig:galaxy_spectra}). A total of two \SI{1200}{\s} exposures were taken on the same night but at two different central frequencies, \SI{6650}{\angstrom} and \SI{6750}{\angstrom}, to have coverage in the GMOS-N detector chip gap, with $2\times2$ binning, providing a spatial scale of \SI{0.00292}{\per\pixel} and an instrumental resolution of \SI{4.66}{\angstrom}, sampled at \SI{1.48}{\angstrom\per\pixel}. The seeing conditions were very good during the observation night, with a mean airmass of \num{1.007}.
Fitting Gaussian line profiles to the H$\alpha$ and N~{\small II} lines (rest wavelengths of \SI{6564.6}{\angstrom} and \SI{6585.2}{\angstrom}) yields a redshift of $z = \redshift$. Assuming the Planck 2018 cosmology \cite{2020A&A...641A...6P}, this redshift implies a Galactic extinction- and k-corrected absolute $r$-band magnitude of \num{-22.03 +- 0.02}. 

The spectroscopic redshift of the galaxy (lines shown in Extended Data Fig.~\ref{fig:sed}) implies an angular diameter distance of \SI{639}{\mega\parsec} and a transverse angular distance scale of \SI{3.1}{\kilo\parsec\per\arcsec}. Using these values, we measure a projected spatial offset for the FRB of \SI{7.2}{\kilo\parsec} from the host galactic centre along the host galactic plane.
This offset is consistent with the distribution of projected offsets measured from a sample of both repeating and non-repeating FRBs localized by the Australian SKA Pathfinder (ASKAP, see e.g., Figure~9 in \cite{2022AJ....163...69B}), with the caveat that our localization ellipse is too large to draw any meaningful conclusion about the host offset.

To characterize the host galaxy, we combined the Gemini spectra with archival photometry from the Two Micron All Sky Survey (2MASS) \cite{2006AJ....131.1163S} and the Wide-Field Infrared Space Explorer (WISE) \cite{2010AJ....140.1868W} to extend our wavelength coverage upwards to \SI{1e5}{\angstrom} (see Methods: Host Galaxy Analysis).

We fit a spectral energy distribution (SED) model to the combined spectral and photometric data using the Bayesian SED-fitting package \texttt{Prospector} \cite{2021ApJS..254...22J}. We estimate best-fit values and uncertainties for the present-day stellar mass, mass-weighted age, V-band dust extinction, and metallicity of our host galaxy using Markov-Chain Monte Carlo (MCMC) posterior sampling (Extended Data Fig.~\ref{fig:sed}; \cite{2013PASP..125..306F}). The parameters determined by \texttt{Prospector} and the star formation rate (SFR) are shown in Table~\ref{table:frbparams}. From the H$\alpha$ luminosity measured with Gemini data, we determine the galaxy's overall SFR (\SI[parse-numbers=false]{0.24 \pm 0.06 \Msun}{\per\year})
and detect star formation in the $\sim 10$ kiloparsec-scale vicinity of the FRB. The detection of H$\alpha$ emission is potentially a sign of recent (\SI{\sim10}{\mega\year}) star formation and young stellar populations.
However, as with the case of other FRBs, spatially-resolved spectroscopic studies of this galaxy are needed to further constrain the age and nature of the FRB progenitor.

In addition to the host galaxy properties, the burst itself can provide insight into the sightline toward the FRB progenitor and the progenitor itself. For instance, if the FRB is located in the inner disk, it would experience enhanced dispersion and scattering due to the long line-of-sight path out of the host galaxy's ionized disk towards the observer, similar to pulsars at low Galactic latitudes in the Milky Way. FRB 20210603A therefore allows for a detailed accounting of host-galactic contributions to the observed DM, RM, and scattering (i.e. pulse broadening). To check this possibility, we calculated the DM excess by subtracting estimated DM contributions from the Milky Way, the Milky Way halo, and the intergalactic medium (IGM) from the measured DM. We obtain a large DM excess of $\text{DM}^\text{r}_\text{host} = \dmhostbudgetrest$, where the superscript denotes that $\text{DM}^\text{r}_\text{host}$ is defined in the host galaxy's rest frame. 

One interpretation of this excess is that of a dense environment local to the FRB progenitor~\cite{2022Natur.606..873N}, which may add significant contributions to the DM, RM, and/or scattering timescale. Another interpretation is that the host galaxy itself contributes a significant portion of the DM excess, with subdominant circumburst contributions to the other properties. Our estimate of the DM budget of the host galaxy is $\sim \dmhostestimaterest$ (see Methods: Dispersion and Scattering Analysis) and is consistent with the latter hypothesis. While both interpretations are compatible with the data in hand, Occam's razor leads us to favor the interpretation that the excess DM of this FRB is dominated by the host galaxy's disk (see Extended Data Fig.~\ref{fig:dm_rm_dist_scatt}). 

This is consistent with our measurements of the pulse broadening timescale, which we determine by fitting a pulse model to the FRB's dynamic spectrum. The complex time-frequency structure of the bright main burst requires three sub-pulse components, temporally broadened by the same characteristic timescale, to obtain a robust fit to the data (see Methods: Burst Morphology and~\cite{2021ApJS..257...59C}). This places an upper limit on the scattering timescale of $\tau_\text{\SI{600}{\mega\hertz}} \lesssim \SI{165 \pm 3}{\micro\second}$ at a reference frequency of $\SI{600}{\mega\hertz}$. Since the scattering from the Milky Way is expected to be subdominant ($\sim$\tauscmodel)~\cite{2002astro.ph..7156C, 2003astro.ph..1598C}, we conclude that the observed pulse broadening is dominated by unresolved substructure in the burst profile or extragalactic scattering, likely in the host rather than the Milky Way~\cite{masui2015dense}.
If the measured broadening timescale is attributed entirely to scattering and scaled to the rest frame and scattering geometry of the host galaxy, the implied scattering efficiency of the host galactic gas is similar to a typical sight-line toward a pulsar through a galactic disk with Milky Way-like density fluctuations (see Methods: Dispersion and Scattering Analysis). 

In addition, the interpretation of a dominant host galactic contribution is consistent with our measurement of the burst RM (see Methods: Polarisation Analysis). After subtracting Galactic and terrestrial contributions ($\mathrm{RM_{MW}, RM_{iono}}$; see Table~\ref{table:frbparams}), the excess is $\mathrm{RM_{excess}} = \rmexcess$. Since no intervening systems (e.g., galaxy groups/clusters) have yet been observed along this sight-line, the RM contribution from the IGM is likely negligible~\cite{Akahori2016}. The magnitude of the RM excess is unremarkable and easily be explained by contributions from the host galaxy's ISM. These properties suggest that the source of FRB 20210603A is located close to its galactic plane (see Extended Data Fig.~\ref{fig:dm_rm_dist_scatt}), consistent with our localization ellipse.

In conclusion, we have commissioned a VLBI array to demonstrate the first VLBI localization of a non-repeating FRB. The limitations of our ad-hoc VLBI array, however, leads to a final localization uncertainty on par with connected-element interferometers like ASKAP, DSA-110, and MeerKAT. Nevertheless, this paves the way towards precisely localizing a large sample of one-off bursts using VLBI.
The FRB 20210603A sightline has implications for galactic astrophysics and the progenitors of FRBs. It demonstrates the potential for using edge-on FRB host galaxies as probes of the ionized gas of other galaxies. In addition the H$\alpha$ emission in the neighbourhood of the FRB suggests recent star formation activity. This highlights the need for high-resolution follow-up to discriminate among progenitor models by assessing whether FRBs are spatially coincident with star-forming regions \cite{2021ApJ...908L..12T}. The instruments and methods used here constitute pathfinders for the CHIME/FRB Outriggers project, which will enable VLBI localizations of large numbers of both repeating and non-repeating sources \cite{2021AJ....161...81L,2022AJ....163...65C,2022AJ....163...48M}. Thus, a more complete picture of the diverse host environments of FRBs, and how the environments correlate with other burst properties, will soon be available.

\section*{Methods}
\label{sec:methods}

\subsection*{Instrumentation and Observations}
\label{sec:instrumentation}

We use a VLBI network consisting of three stations: the Canadian Hydrogen Intensity Mapping Experiment (CHIME) at the Dominion Radio Astrophysical Observatory (DRAO) \cite{2018ApJ...863...48C}, ARO10, a 10-m single dish at Algonquin Radio Observatory (ARO) \cite{2022AJ....163...65C}, and TONE, a compact array of eight 6-m dishes at Green Bank Observatory (GBO) \cite{tonesystem}. CHIME/FRB detected FRB 20210603A at 2021-06-03~15:51 UTC. In Fig.~\ref{fig:frb_i_stokes} we show the Stokes-I dynamic spectrum of the beamformed data from FRB 20210603A as observed at CHIME. Between August 2018 and May 2021, \SI{35.6}{\hour} of exposure were accumulated in the direction of FRB 20210603A; however only the burst reported here was detected. For VLBI calibration and testing our localization procedure, we used several Crab GPs captured at a cadence of one per day, which we refer to as C1--C4 respectively (see Extended Data Fig.~\ref{fig:i_stokes_events}). 

\subsubsection*{CHIME/FRB}
\label{sec:chime}
CHIME consists of four $\SI{20}{\m}\times\SI{100}{\m}$ cylindrical paraboloid reflectors oriented with the cylinder axis in the North-South direction \cite{2022ApJS..261...29C}. Each cylinder is fitted with \SI{256} dual-linear-polarisation antennas that are sensitive in the frequency range of \SIrange{400}{800}{\mega\hertz}.
The \num{2048} analog signals from the antennas are amplified and digitized using an array of \num{128} field programmable gate array (FPGA) driven motherboards with mezzanine analog-to-digital converters (ADCs) called ICE boards \cite{2016JAI.....541005B}. 
At each ICE board, raw voltages are channelized with a polyphase filterbank (PFB) producing \num{1024} complex channels with \SI{2.56}{\micro\s} time resolution. We refer to the channelized and time-tagged voltage data as raw baseband data (as opposed to beamformed baseband data, see Methods: Local Calibration and Beamforming). 
These data are sent to \num{256} GPU-based compute nodes comprising the X-Engine correlator driven by the open-source \texttt{kotekan} software repository~\cite{2020JAI.....950014D,2021zndo...5842660R}.
Here, the spatial correlation is computed and polarisations are summed, forming \num{1024} ($256\text{-NS}\times4\text{-EW}$) independent beams within the North-South primary beam \cite{2017ursi.confE...4N}. These beams are searched for FRBs in real-time using detection pipelines designed for discovering radio transients.
The real-time pipeline and the baseband system collectively make up the CHIME/FRB instrument \cite{2018ApJ...863...48C,2021ApJ...910..147M}. The baseband system uses a memory ring buffer system to record (or ``dump'') baseband data to disk. The ring buffer holds \SI{\sim35.5}{\s} of baseband data for subsequent capture by a detection trigger. On successful detection of an FRB candidate by the real-time pipeline above an S/N of \num{12}, a trigger from the real-time pipeline saves a \SI{\sim100}{\milli\s} snapshot of data centred around the pulse at each frequency channel of the baseband buffer. The latency between the time of arrival of a signal and the triggered baseband recording is typically \SI{\sim14}{\s}. The buffer can record the full band's worth of data when the dispersive sweep of the FRB does not exceed \SI{\sim20}{\s} (corresponding to a maximum DM of \SI{\sim1000}{\parsec\per\centi\m\cubed}). 

The outrigger triggering system involves asynchronous servers running at ARO10 and TONE. Each station sends a ``heartbeat'' to the CHIME/FRB backend. The CHIME/FRB backend then registers each outrigger with a heartbeat as an active outrigger. Upon detection by the real-time pipeline of an FRB or a Crab pulsar GP \cite{2015MNRAS.446..857L} in the FoV of TONE and ARO10, a trigger is sent to the active outriggers. To prevent GP triggers overwhelming the baseband readout system with thousands of events, we record only triggers with a detection S/N greater than \num{40} (near CHIME's zenith) having a duty cycle of \SI{1}{\percent}. This results in a Crab GP dump rate of about once per day.

\subsubsection*{Algonquin Radio Observatory 10-m telescope}
\label{sec:aro10m}

ARO10, a 10-m single dish, is located at the Algonquin Radio Observatory in Algonquin Provincial Park, Ontario. The CHIME-ARO10 baseline is over $b_{CA} \gtrsim \SI{3000}{\kilo\m}$ (see Figure~\ref{fig:vlbi_projection}). The two analog signals from the polarisations of the single CHIME cloverleaf feed \cite{2017arXiv170808521D} are digitized and acquired with a digital infrastructure identical to that of CHIME and TONE except that the large (\SI{\sim24}{\hour} long) ring buffer is stored on hard disks. A complete description of the radio frequency (RF) chain and the digital system is provided elsewhere \cite{2022AJ....163...65C}. The data at ARO10 exhibit a delay drift relative to DRAO amounting to \SI{\sim0.1}{\micro\s\per\day}. This extra shift in addition to the \SI{\sim2}{\milli\s} geometrical delay is predictable and is corrected (see Figure~15 of~\cite{2022AJ....163...65C}).

\subsubsection*{TONE}
\label{sec:tone}

TONE is located at GBO near the Green Bank Interferometer Control Building. The CHIME-TONE baseline is $b_\mathrm{CT}\approx\SI{3332}{\kilo\m}$ long (see Figure~\ref{fig:vlbi_projection}). TONE is an array of 6-m dishes placed in a regular $4\times3$ grid with 9.1-m spacing with the shorter side aligned \SI{60}{\degree} off true north. Each dish is oriented to observe the Crab pulsar at the same time as CHIME. Eight dishes are deployed with feeds instrumented with active-balun dual-polarised cloverleaf antennas \cite{2017arXiv170808521D,2022JATIS...8a1019C}. The \num{16} analog signals are each transmitted over a radio-frequency-over-fiber (RFoF) system \cite{rfof}. For this work, \num{7} signals from one polarisation and \num{6} signals from the other were used to synthesize a single beam for VLBI. The signals from the RFoF receiver are digitized and channelized by an ICE board (in the same way that was previously described for CHIME and ARO10). A TM-4 GPS clock module \cite{tm-4} provides a \SI{10}{\mega\hertz} clock and absolute time. Additionally, a \SI{10}{\mega\hertz} maser signal is fed into the ICE board \cite{2022AJ....163...48M} replacing one of the analog inputs for post-hoc clock delay characterization \cite{2021RNAAS...5..216C}. The digitized and channelized voltages are sent via two \SI{40}{\giga\bit} ethernet network links over to the recording computer node. The recording node uses \texttt{kotekan}, as it does at CHIME and ARO10, to create a \SI{\sim40}{\second} buffer of the baseband data~\cite{2021AJ....161...81L}. The length of the buffer must accommodate both the latency of the CHIME/FRB detection pipeline and the network in addition to the science data. The baseband readout saves a \SI{\sim 0.5}{\second} slice of the buffer around the pulse on the arrival of a trigger to disk for offline VLBI analysis. Taurus A is used as a calibrator to phase the antennas within TONE for beamforming (see Methods: Local Calibration and Beamforming). 
See \cite{Sanghavi2022frb} for a detailed description of the system and its performance.

We have summarized in Table \ref{tab:telescope_properties} the three sites and their properties. Further description of the systems and correlation routine can be found in \cite{cassanelli2022frb,leung2023frb,Sanghavi2022frb}.

\subsection*{Clock Calibration}
\label{clock}
There exist timing errors intrinsic to the digital backends at each station, which are locked to different clocks with varying degrees of stability. The severity of timing errors depends on the type of clock used at each station and varies from unit to unit. Timing errors are characterized in terms of the Allan deviation ($\sigma(\Delta t)$) as a function of timescale $\Delta t$, e.g. between successive clock calibrations \cite{2022AJ....163...48M}. 
The CHIME digital system is locked to a single \SI{10}{\mega\hertz} clock signal provided by a GPS-disciplined, oven-controlled crystal oscillator. While sufficient for the operations of CHIME as a stand-alone telescope, this clock does not meet the stringent stability requirements for VLBI with CHIME/FRB Outriggers. To overcome this limitation, we sample the more stable passive hydrogen maser (located at the DRAO site) during FRB VLBI observations \cite{2022AJ....163...48M} on a regular cadence. This minimally-invasive clocking system was developed as part of the effort to expand CHIME's capabilities to include VLBI with CHIME/FRB Outriggers. 
It works by digitizing the signal from an external maser using one of the inputs of the GPS-clock-driven ICE board. We read out a \SI{2.56}{\micro\second} snapshot of maser data at a cadence of once every $\Delta t_{\text{GPS,C}} =\SI{30}{\second}$ at CHIME. The data readout from the maser are processed offline to measure the drift of the GPS clock between calibrator observations.
A similar readout system records a \SI{10}{\mega\hertz} clock at TONE at a cadence of $\Delta t_{\text{GPS,T}} =\SI{1}{\second}$. In contrast, the digital system of ARO10 is directly clocked by an actively-stabilized hydrogen maser, removing the need for station-based clock corrections. 

Once clock corrections are applied to the observations, the expected delay error between two observations separated by $\Delta t_{sep}$ in time is given by the quadrature sum of the jitter at each station. Assuming that the jitter is characterized by the Allan deviation of the maser alone, this is given by $\sigma_{\text{maser}}(\Delta t_{sep}) \Delta t_{sep}$. On 24-hour timescales, this corresponds to a delay error of $\approx \SI{0.35}{\nano\second}$ for the CHIME-ARO10 baseline (one passive,
one active maser), and $\approx \SI{0.48}{\nano\second}$ for the CHIME-TONE baseline (two passive masers)~\cite{2022AJ....163...48M}. In addition, on the CHIME-TONE baseline, observations are referenced to the maser by interpolating between the maser readouts directly before and after the observation. The slow cadence of maser readout at these stations induces an additional interpolation error of size $\sigma_{\text{GPS}}(\Delta t_{\text{sync}}) \times \Delta
t_{\text{sync}}$~\cite{2021RNAAS...5..216C}, for a total of $\SI{0.52}{\nano\second}$.

\subsection*{Local Calibration and Beamforming}
\label{sec:beamforming}

CHIME has \num{1024} antennas, and TONE has \num{8} antennas. It is infeasible to correlate such a large number of antennas as independent VLBI stations. To reduce the computational burden of correlating such a large array, we coherently add, or beamform, the raw baseband data from the antennas within each station to combine the multiple low-sensitivity antennas from a single station into a high-sensitivity equivalent single dish using beamforming. 

Beamforming requires independent measurements of the individual sensitivities and delays for each antenna, i.e., complex-valued gains which contain both amplitude and phase information. At CHIME, the infrastructure to calculate these so-called ``$N^2$-gains'' and a tied-array beamformer have already been developed \cite{2022ApJS..261...29C}. We generalized several of CHIME's software frameworks \cite{2015arXiv150306189R,2014SPIE.9145E..22B}, to use the same basic $N^2$-gain calibration algorithms \cite{2014SPIE.9145E..4VN} at TONE. First, the visibility matrix from all $N^2$ pairs of antennas at the correlator is calculated when a bright point source (Taurus A for TONE) dominates the FoV. In the single-source limit, the visibility matrix has a rank-\num{1} eigendecomposition; the non-singular eigenvector and eigenvalue encode a combination of geometric delays and instrumental gains and delays. 
Once the gains are characterized, they are used to beamform the raw baseband data from CHIME and TONE towards the best-known positions of the Crab and the FRB provided by the baseband pipeline ($\widehat{\mathbf{n}}_0$). The synthesized beam at CHIME is \SI{\sim1}{\arcmin} wide, and the synthesized beam at TONE is \SI{\sim0.5}{\degree} wide. Since the FRB's true position is well within a synthesized-beam width away from $\widehat{\mathbf n}_0$, our final sensitivity only depends weakly on $\widehat{\mathbf n}_0$. 

\subsection*{VLBI Correlation}
\label{sec:localization_analysis}

After beamforming is completed at each station, the beamformed baseband data are correlated with a custom Python-based VLBI correlator~\cite{2024arXiv240305631L}. 
We use the standalone delay model implemented in \texttt{difxcalc}~\cite{2016ivs..conf..187G} to calculate geometric delays towards the fiducial sky location $\widehat{\mathbf{n}}_0$ of each source. For the Crab pulses, we use the VLBI position of the Crab pulsar~\cite{crabvlbi} extrapolated using its proper motion to the epoch of our observations:
\begin{equation}
    \widehat{\mathbf{n}}_0 = \del{\SI{83.6330379}{\degree}, \SI{22.014501}{\degree}},
\end{equation}
with RA and Dec components reported in decimal degrees.
Including the pulsar position error ($\sigma_{\widehat{\mathbf{n}}}$) and the proper motion ($\boldsymbol{\mu}$) error ($\sigma_{\boldsymbol{\mu}}$) extrapolated over $\approx \SI{10}{\year}$ from recent Crab pulsar astrometry~\cite{crabvlbi}, we sum the absolute position error at the archival observing epoch and the uncertainty in the proper motion, scaled by the time between our observations (\SI{\sim10}{\year}), in quadrature for the RA and Dec components individually. The uncertainties in the Crab position propagate into equally-sized positional uncertainties of the FRB; however these are subdominant compared to our systematics, so we do not quote them above. For the FRB, we use the best-fit position derived from a CHIME-only baseband localization ($\widehat{\mathbf{n}}_0 = (\SI{10.2717}{\degree}, \SI{21.226}{\degree})$). This is precise to within an arcminute; nevertheless, we find strong fringes on the FRB pointing towards this position. 

In our correlator, we break the total delay into an integer number of \SI{2.56}{\micro\second} frames and a sub-frame (or sub-integer) component whose value is in the range \SIrange{-1.28}{1.28}{\micro\second}. The integer shift is applied to the data via an array shift, and the sub-integer shift is applied by a phase rotation to each $\SI{2.56}{\micro\second}$ frame. While this time resolution is lower than that of more conventional VLBI backends, doing delay compensation on this timescale does not appreciably increase phase errors, even at the top of the band where these would be most noticeable. We estimate an upper limit on the phase error at the top of our band to be $\sim \epsilon \times \SI{2.56}{\micro\second} \times \SI{800}{\mega\hertz}$, where $\epsilon$ is the maximum delay rate encountered during our observations. For the most extreme scenario of two antipodal VLBI stations located on the equator, $\epsilon \approx 3\times 10^{-6}$ gives a phase error of \SI{2.2}{\degree}: an acceptably small amount of decorrelation.

After delay compensation, each of the \num{1024} frequency channels of data is de-smeared by a coherent dedispersion kernel \cite{1975MComP..14...55H}. While several conventions may be used (see e.g., Eq.~5.17 in \cite{2012hpa..book.....L}), we use the following kernel in our VLBI correlator:
\begin{equation}
    H(\nu) = \exp \left( 2\pi \mathrm{i} k_{\mathrm{DM}} {\mathrm{DM}} \dfrac{\nu^2 }{2\nu_k^2 (\nu_k + \nu)} \right).
    \label{eq:kernel}
\end{equation}
In Eq.~\eqref{eq:kernel}, we take  $k_\text{DM}=\num{1}/(\num{2.41e-4})$ \si{\s\mega\hertz\squared\per\parsec\centi\m\cubed} (for consistency with previous conventions in the pulsar community \cite{2012hpa..book.....L,2020arXiv200702886K}), and the fiducial DM of the FRB is taken to be \dmfrb. We choose this dedispersion kernel in order to avoid introducing delays into each frequency channel (i.e. it preserves times of arrival at the central frequency of each channel). The chosen DM de-smears the pulse within each frequency channel. This concentrates the signal into a narrow temporal duration and increases the correlation power. The argument $\nu \in \sbr{-\SI{195.3125}{\kilo\hertz},+\SI{195.3125}{\kilo\hertz}} $ indicates the offset from the reference $\nu_k$, chosen to be the centre of each frequency channel: $\nu_k \in \sbr{\num{800.}, \num{799.609375},...,\num{400.390625}}$ \si{{\mega\hertz}}. 

After the delay compensation towards the fiducial sky position $\widehat{\mathbf{n}}_0=\del{\upalpha_0,\updelta_0}$ and coherent dedispersion, we form visibilities for each frequency channel (indexed by $k$) independently on both long baselines involving CHIME ($b_\mathrm{CA}$ and $b_\mathrm{CT}$, hereafter indexed by $i$) by multiplying and integrating the complex baseband data. To reject noise, we integrate only $\SI{\sim100}{\micro\second}$ of data on either side of the pulse in each of \num{1024} frequency channels. In addition, we rejected RFI channels (see Extended Data Fig.~\ref{fig:i_stokes_events}) within each site. This produces \num{\sim900} complex visibilities per baseline which are used for localization (hereafter referred to as $V\sbr{i,k}$).  We integrate \num{13} other windows of the same duration in the same dataset but shifted to off-pulse times to estimate the statistical uncertainties on the visibilities. The statistical uncertainties are hereafter referred to as $\sigma\sbr{i,k}$.

\subsection*{VLBI Calibration and Empirical Localization Error Budget}
The complex visibilities $V\sbr{i,k}$ must be phase-calibrated prior to the localization analysis. We calibrate the visibilities with phase, delay, and rate corrections derived from our Crab GPs before performing our final localization analysis. In an ideal setup, we would systematically characterize localization errors in the CHIME-ARO10-TONE array as a function of sky pointing and time separation and perform end-to-end localization of known pulsars as a checks of our localization. However, our ability to do so is limited due to logistical factors at each station. Perhaps most logistically difficult is the extremely limited internet access to the ARO10 site, which fundamentally limits the data that can practically be read out from the ARO10 site \cite{2022AJ....163...65C}. At TONE, frequent misalignment of the dishes due to high wind conditions requires manual repointing and recalibration of the array, which frequently interrupts observations. Therefore, the only data available for characterizing the full CHIME-ARO10-TONE array around the time the FRB was observed are a sequence of triggered baseband dumps from the Crab pulsar collected in May--June 2021, simultaneous with CHIME, occurring at a cadence of about \num{1} per day, at each station. We enumerate these Crab pulses as C1--C4. Waterfall plots of these pulses, in addition to the FRB, are shown in Extended Data Fig.~\ref{fig:i_stokes_events}.

Within the constraints of these limited data, we perform the following steps for VLBI calibration. We use C2, the closest Crab pulse in time to the FRB, as a delay and phase calibrator, i.e. we calculate instrumental phase and delay solutions for all baselines, and apply them to all observations on all baselines. The phase and delay solutions remove static instrumental cable delays and frequency-dependent beam phases, and suppress unwanted astrometric shifts related to baseline offsets towards the elevation angle of the Crab, which is less than a degree away from the FRB in alt-azimuth coordinates. In addition to the phase and delay calibration, a large delay rate correction (\SI{\sim0.1}{\micro\second\per\day}) is needed for the CHIME-ARO10 baseline~\cite{2022AJ....163...65C}. Upon removal of the CHIME-ARO10 clock rate, our delay residuals are small (see Extended Data Fig.~\ref{fig:empirical_error}). In that Figure we also include all of the delay residuals from historical data available on each baseline individually, calibrated similarly (i.e., with a clock rate correction for CHIME-ARO10 and with no significant clock rate correction detected for the CHIME-TONE). 

In the absence of commissioning data available when all three stations were operating, we characterize each baseline individually.
For CHIME-ARO10, we show a previously-published dataset of \num{10} correlated Crab pulses from October 2020. For CHIME-TONE data, we use \num{11} Crab GPs from the February--March 2021 period during which the instrument was commissioned \cite{tonesystem}. From these data, we establish $1\sigma$ systematic localization uncertainties by calculating the RMS delay errors on each baseline using most of the data plotted in Extended Data Fig.~\ref{fig:empirical_error}. The RMS delay error on the CHIME-ARO10 and the CHIME-TONE baselines are \SI{8.5}{\nano\second} and \SI{6.0}{\nano\second} respectively, calculated from 10 and 11 Crab single-baseline measurements respectively. These RMS values have been calculated excluding the pulses used for delay/rate calibration (whose delay residuals are zero by definition) and faint pulses (CHIME-TONE data from March 2021) whose fringe detections are marginal, due to a windstorm at Green Bank which blew several TONE dishes off-axis before they were manually repointed.

\subsection*{Crab Localisation}
In addition to quantifying delay errors on each baseline individually using Crab pulses, we perform an independent, end-to-end cross-check of the delay and rate solutions derived for the FRB using C3. This is the only Crab GP remaining which is detected at all stations and baselines which we have not used to obtain delay and rate solutions; we use it here as an independent check of our delay and rate solutions and of our localization procedure, which combines data from both baselines.

To localize C3, we calibrate C3 visibilities for both baselines using the aforementioned delay and phase solutions from C2. In addition, on the CHIME-ARO10 baseline we apply the clock rate measured from C1 and C2. The calibrated residual delay when the C3 data are correlated towards the true Crab position is \SI{2.8}{\nano\second} for the CHIME ARO10 baseline and \SI{2.1}{\nano\second} for the CHIME TONE baseline. To further model the short-term trend seen in the CHIME-TONE delay residuals, we attempted to apply a clock rate correction to CHIME-TONE data measured from C2 and C4 (since the TONE correlator restarted between C1 and C2). Doing so only changes the CHIME-TONE delay by \SI{\sim1}{\nano\second}. The residual delays, as well as the final delay rate correction, are subdominant to our $1\sigma$ systematic error budget of 8.5 and \SI{6.0}{\nano\second} for the CHIME-ARO10 and CHIME-TONE baselines.

We refer to the visibilities calibrated this way as $\mathcal{V}\sbr{i,k}$ (not to be confused with the un-calibrated visibilities $V\sbr{i,k}$), where $i$ denotes the baseline (either CA or CT) and $k$ denotes our \num{1024} independent frequency channels. They are plotted with residual delays removed in Extended Data Fig.~\ref{fig:visibilities_crab}.
In addition to the correlation start times in each channel $t_0\sbr{i,k}$, and the baseline vectors $\mathbf{b}_\mathrm{CA}$, $\mathbf{b}_\mathrm{CT}$, we use $\mathcal{V}\sbr{i,k}$ to localize C3 to an inferred position $\widehat{\mathbf{n}}$ relative to the fiducial sky position ($\widehat{\mathbf{n}}_0$) used to correlate C3.

Several approaches to localizing single pulses been taken in the literature~\cite{2021AJ....161...81L,2022AJ....163...65C} and \cite{2022ApJ...927L...3N}, reflecting the significant challenge of astrometry with sparse $uv$-coverage. For example, the traditional method of making a dirty map of a small field and using traditional aperture synthesis algorithms to de-convolve the PSF is not well-suited to the present VLBI network because of the sparse $uv$-coverage.
We have found that one robust method is to take the delay estimated from the peak of the Fourier transform of the visibilities, and use that delay measurement to localize the FRB by maximizing Eq.~\ref{eq:logl_tau}. This method is robust is the sense that Eq.~\ref{eq:logl_tau} only has one global maximum, so it works well even when the true position is arcminutes away from the true position.
\begin{equation}
    \log\mathcal{L}_\tau = \sum_{i=\text{CA,CT}}\dfrac{\del{\tau_i^\mathrm{max} - \tau_i\del{\widehat{\mathbf{n}}}}^2}{2\sigma_{\tau,i}^2}\label{eq:logl_tau}
\end{equation}
The drawback of this simple method is that it only is sensitive to the information contained in the linear part of the phase model ($\dif\phi/\dif\nu_k$), which means it mixes the ionosphere and geometric delays and therefore is only accurate at the arcsecond level. Working in visibility space is a straightforward way to break this degeneracy, since we can fit higher-order contributions to the phase as a function of frequency. We fit Eq.~\ref{eq:phase_model} to our data to disentangle the ionosphere
from the geometric delays:
\begin{equation}
    \phi\sbr{i,k} = 2\pi \del{\nu_k \tau_i + k_{\mathrm{DM}} \Delta\mathrm{DM}_i \frac{1}{\nu_k}}. \label{eq:phase_model}
\end{equation}

We obtain the best fit solution by maximizing the visibility-space likelihood function (Eq.~\ref{eq:likelihood}). Practically, it is difficult to do this because the posterior is highly multimodal as seen in our final contours, which are shown in Extended Data Figure~\ref{fig:mcmc_corner_c3}. We resort to using a box centred on a good initial guess. For the R.A. and declination, the initial guess is taken from the $\mathcal{L}_\tau$ localization. The initial guesses for Delta DM on each baseline were determined by independently optimizing the signal to noise (Eq.~\ref{eq:snr}) over a range of Delta DM and delay values on each baseline.
\begin{equation}\label{eq:snr}
    \rho_\mathrm{sf}(\tau,\Delta\mathrm{DM}) = \enVert{\sum_{k} \frac{\mathcal{V}\sbr{i,k} \exp\del{-\mathrm{i}\phi\sbr{i,k}}}{\sigma\sbr{i,k}}}
\end{equation}

With these initial guesses we evaluate Eq.~\ref{eq:likelihood} on a 4D grid to simultaneously solve for the source position and the ionosphere parameters. Eq.~\ref{eq:likelihood} uses a signal-to-noise weighting scheme, weighting the real part of the phase-rotated visibilities by $|V| / \sigma^2$. The denominator of this weighting corresponds to inverse noise weighting; $\sigma\sbr{i,k}$ refers to the statistical uncertainties in the visibilities. The numerator corresponds to an upweighting
by the visibility amplitude. Since the FRB is detected in each channel with a signal-to-noise of $\sim 5-10$, and since it is the single dominant source of correlated flux in the correlated data, we use the visibility amplitude $|V\sbr{i,k}|$ as a convenient approximation to the statistically-optimal upweighting, which is the true signal power in each channel after applying appropriate bandpass and beam corrections to each baseline. Note that the band-integrated signal-to-noise reported elsewhere (e.g. Extended Data Fig.~\ref{fig:visibilities_crab}) is an underestimate of the true signal-to-noise, since the flux from the FRB contributes significantly to the RMS noise level of the FFT.

\begin{equation}
\log \mathcal{L}_\varphi \propto \sum_{i=\text{CA,CT}}\sum_{k=0}^{1023} \dfrac{\enVert{\mathcal{V}[i,k]}\mathrm{Re}\sbr{\mathcal{V}\sbr{i,k}\exp\del{-\mathrm{i}\phi\sbr{i,k}}}}{\sigma\sbr{i,k}^2.}
\label{eq:likelihood}
\end{equation}

The posterior as a function of our four parameters $(\alpha,\delta,\Delta\mathrm{DM_{CA}},\Delta\mathrm{DM_{CT}})$ is shown in Extended Data Fig.~\ref{fig:mcmc_corner_c3}. We take the parameter set that maximizes the likelihood on the grid as the best-fit model. The model phases corresponding to these parameters, as well as the model phases corresponding to the parameters which maximize $\mathcal{L}_\tau$  are plotted in Extended Data Fig.~\ref{fig:visibilities_crab}. The
maximum-$\mathcal{L}_\varphi$ position of C3 is $\widehat{\mathbf{n}} = \del{\SI{83.633053}{\degree}, \SI{22.014539}{\degree}}$. Finally, we draw systematic error contours around this best-fit position using $\sigma_{\tau,i} = 8.6,\SI{6.0}{\nano\second}$ respectively in Extended Data Fig.~\ref{fig:c3_loc}. The 1-sigma systematic error contour drawn around the best fit position easily encloses the Crab's true position and the delay-only best-fit position Extended Data Fig.~\ref{fig:c3_loc} which does not separate
out the ionospheric delay, showing that the ionosphere is not the dominant source of systematic error in our localization.

\subsection*{FRB Localization}
We apply the exact same calibration solutions used to localize C3 to the FRB visibilities. Following the same procedure, we use the coarse localization with $\mathcal{L}_\tau$ to coarsely localize the FRB. The $\mathcal{L}_\tau$ position is $\widehat{\mathbf{n}} = (\SI{10.274056}{\degree}, \SI{21.22624}{\degree})$, and is offset from the baseband localization by $ \SI{8}{\arcsec}$ in the RA direction and $\approx\SI{-1.3}{\arcsec}$ in the declination direction. To
recover some sensitivity, we re-point the correlator phase center towards this refined position before fringe fitting the calibrated visibilities (Extended Data Fig.~\ref{fig:visibilities_frb}) for the ionosphere using $\mathcal{L}_\varphi$. The initial guesses for the ionosphere are estimated as done previously for C3, and the fringe fit yields the maximum-likelihood position $\widehat{\mathbf{n}} = \del{\radecimal, \decdecimal}$ (Table \ref{table:frbparams}). The posteriors are shown in Extended Data Fig.~\ref{fig:mcmc_corner_frb}.

\subsubsection*{Possible Error Sources} 
\label{sec: error}
We summarize some known contributions to our systematic error, which we find cannot account for the empirically measured delay errors (\num{8.5} and \SI{6.0}{\nano\second} at $1\sigma$). We have seen that this corresponds to a $\ang{;;0.2} \times \ang{;;2}$ ellipse on the sky, and that relative to this ellipse, the effect of including the ionosphere is small. We estimate our station positioning errors to be \SI{21}{\mas} assuming a conservative $\approx \SI{10}{\meter}$ baseline uncertainty. Time
variations in the phasing of the antennas may also arise at CHIME or TONE, since relative cable lengths fluctuate on weeklong timescales by around $\SI{0.1}{\nano\second}$ at these stations, but they are also re-calibrated every day. Uncertainties in the proper-motion extrapolated position of the Crab pulsar at its current epoch (\SI{2}{\mas}) are also subdominant. Another systematic uncertainty is the astrometric frame tie between our VLBI localization (ICRF) and optical follow-up observations, which are performed relative to the FK5/ICRS reference frame. The discrepancy between the frames is on the order of $\sim\SI{1}{\mas}$~\cite{1998AJ....116..516M,2021ARA&A..59...59B,2005USNOC.179.....K,1998A&A...331L..33F}.

Since none of these explain the magnitude of our systematic error, we have to consider alternate sources of delay fluctuations. One-day timescale variations in the masers' relative oscillation frequencies or the signal chains carrying the maser signals to the digitizers in the F-engine may add delay noise on timescales relevant for our sparse calibration. Regardless, our empirical measurement of the RMS delay residuals (see Extended Data Fig.~\ref{fig:empirical_error}) to quantify our localization uncertainty encompasses all of the known and unknown sources of
systematic astrometric uncertainties, putting our scientific conclusions on firm footing. In the future, dedicated lab tests could verify this. To avoid the issue completely, the time between VLBI calibrations could also be shortened to minutes or hours. With future outrigger stations having significantly more collecting area than ARO10 and TONE, this will be readily achievable.

\subsection*{Burst Morphology}
FRB 20210603A was detected with a signal-to-noise ratio of \num{\sim 136} in the CHIME/FRB real-time detection pipeline. Afterwards, we characterized its burst morphology and estimated its brightness using high-resolution baseband data; the flux, fluence, specific energy, and specific luminosity of the burst are listed in Table \ref{table:frbparams}. Viewed in baseband data, the FRB has a broadband main pulse with a total full-width half maximum of \SI{740}{\micro\second}. In addition, two trailing components are visible in the baseband dump (Fig.~\ref{fig:frb_i_stokes}). Using the \texttt{DM\_phase} algorithm~\cite{2019ascl.soft10004S}, we line up substructures in the main pulse, yielding a DM of \dmfrb. The DM and the baseband data are inputted to \texttt{fitburst} \cite{2024ApJS..271...49F}, which simultaneously fits the main burst with three closely-spaced sub-bursts with full-width half maximum widths of 310, 450, and \SI{834}{\micro\second}, all broadened by \SI{165}{\micro\second} at \SI{600}{\mega\hertz}. 

\subsection*{Dispersion and Scattering Analysis}
In general the observed DM of an FRB can be split into four components as, 
\begin{equation} \label{eq:dmbudget}
    \mathrm{{DM}_{FRB}} = \mathrm{{DM}_\text{MW-disk}} + \mathrm{{DM}_\text{MW-halo}} + \mathrm{{DM}_{cosmic}} + \mathrm{{DM}_{host}}, 
\end{equation}
where  $\text{DM}_{\text{MW-disk}}$ is the contribution of the disk of the Milky Way, $\text{DM}_{\text{MW-halo}}$ is that from the extended hot Galactic halo and $\text{DM}_{\text{cosmic}}$ is from the intergalactic medium. The DM contribution of the host, $\text{DM}_{\text{host}}$, is a combination of the contributions from the interstellar medium (ISM) of the host galaxy $\text{DM}_{\text{host-disk}}$, the halo of the host galaxy  $\text{DM}_{\text{host-halo}}$ and the contributions from the source environment $\text{DM}_{\text{host-env}}$.

To interpret unknown contributions to the total DM, we subtract known contributions from the total. To estimate the contribution from the Milky Way disk we default to the NE2001 model~\cite{2002astro.ph..7156C,2003astro.ph..1598C} obtaining $\text{DM}_{\text{MW-disk,NE2001}} = \dmmwne$, noting that the YMW16 model~\cite{ymw16} yields similar results. We estimate the contribution of the Galactic halo to be $ \text{DM}_{\text{MW-halo}} = \dmmwhalo$ using the model described in \cite{2020ApJ...888..105Y}. We can treat this estimate as conservative, and it can be as low as \SI{6}{\parsec\per\centi\meter
\cubed}~\cite{2020MNRAS.496L.106K}. This is also consistent with CHIME/FRB constraints on the halo DM~\cite{2023ApJ...946...58C}.
The IGM contribution is estimated to be $\text{DM}_\text{cosmic} = \dmcosmic$~\cite{2020Natur.581..391M}, where the range is due to cosmic variance in the Macquart relation out to $z \approx \num{0.18}$~\cite{2021MNRAS.505.5356B}. This leaves the contribution to the DM from the host galaxy halo, disk, and the FRB local environment as $\text{DM}_\text{host}$ = \dmhostbudget. 

The large value of $\text{DM}_\text{host}$ is consistent with a long line-of-sight traveled through the host galaxy disk, resulting from the galaxy inclination angle. We can estimate the DM contributions of the host galaxy disk and halo by scaling the Milky Way's properties using the stellar mass of the host galaxy (see Methods: Host Galaxy Analysis). We assume the disk size ($R$) scales with the galaxy stellar mass $\rm{M}_\text{host}^\star$ as a power law $R \propto \del{\rm{M}_\text{host}^\star}^\beta$ where for simplicity we choose $\beta \sim 1/3$. This value of $\beta$ is close to the measured value in the literature for galaxies with $\text{M}^\star = \numrange[retain-unity-mantissa=false]{1e7}{1e11}\Msun$ \cite{2020MNRAS.493...87T}. Thus the galaxy size scales as $\del{\text{M}_\text{host}^\star/\text{M}^\star_\text{MW}}^{1/3} = (\massratio)^{1/3} = \massscalingfactor$, where $\text{M}_\text{host}^\star = (\masshost) \tentoten \Msun$ and $\text{M}^\star_\text{MW} = (\massmw) \tentoten \Msun$ are the present-day stellar masses of the Milky Way \cite{2015ApJ...806...96L} and the host galaxy respectively,. Assuming the halo size also scales as $\del{\text{M}_\text{host}^\star/\text{M}^\star_\text{MW}}^{1/3}$, the average Milky Way halo DM contribution \dmmwhaloavg \cite{2020ApJ...888..105Y} can be scaled to estimate  $\text{DM}^\text{r}_\text{host-halo} = \text{DM}_{\text{MW-halo}} \times (\text{M}_\text{host}^\star/\text{M}^\star_\text{MW})^{1/3} = \dmhosthalorest$ in the host galaxy's rest frame. 
Similarly, we can conservatively estimate the rest frame DM due to the disk of the host galaxy, $\text{DM}^\text{r}_\text{host-disk}$. A first approximation is to assume that the FRB originates from close to the midplane of the disk, and scale the DM contribution of the half-thickness of the Milky Way ($N_{\perp}(\infty) \approx \dmmwperp$ \cite{2020ApJ...897..124O}) by a factor of $\csc\del{\inclination} = \geometryfactor$ to account for the viewing geometry. 
We assume the electron density stays equivalent to that of the Milky Way and scale for the host galaxy size. This yields an estimate of $\text{DM}^\text{r}_\text{host-disk} = N_{\perp}(\infty) \times \csc\del{\inclination} \times \del{\text{M}_\text{host}^\star/\text{M}^\star_\text{MW}}^{1/3} = \dmhostdiskrest$ in the host galaxy rest frame. We can sum these estimates of the $\text{DM}^\text{r}_\text{host-disk}$ and $\text{DM}^\text{r}_\text{host-halo}$ to give the DM in the observer's frame as $\text{DM}_\text{host} = (\text{DM}^\text{r}_\text{host-disk} + \text{DM}^\text{r}_\text{host-halo})/(1+z) = \dmhostestimate$ which is consistent with the observed $\text{DM}_\text{host}$. If the FRB is behind the galaxy, the expected contribution from the host galactic disk could be increased by up to a factor of \num{2} yielding \dmhostestimatebehind; however, this possibility is inconsistent with the observed DM excess.

In addition to the DM of the host galaxy, we can also measure gas fluctuations in the host galaxy using pulse broadening. The measured pulse broadening timescale from \texttt{fitburst} is $\tau_\text{scatt-\SI{600}{\mega\hertz}} = \SI{165 \pm 3}{\micro\second}$. However on visual inspection of the dynamic spectrum, we cannot rule out the possibility that this timescale originates from unresolved downward-drifting substructure. We treat this timescale as an upper limit on the true scattering timescale, and consider the implications for the FRB progenitors and the host galactic gas by comparing the dispersion and scattering to Galactic pulsars at similar Galactic latitudes. To place these measurements on equal footing, we scale $\tau_\text{scatt-\SI{600}{\mega\hertz}}$ to $\SI{1}{\giga\hertz}$, and multiply by $(1+z)^3$ to account for time dilation and the un-redshifted frequency at which the pulse is scattered. This gives $\tau_\text{proper,\SI{1}{\giga\hertz}} = \SI{45}{\micro\second}$ in the rest frame of the host galaxy. Further dividing this by \num{3} converts the geometric weighting from that of extragalactic (plane-wave) scattering to Galactic (spherical-wave) scattering \cite{2021ApJ...911..102O}. Finally, subtracting $\mathrm{DM}_\text{host-halo}$ from the observed DM excess in the host galaxy rest frame yields $\mathrm{DM}^\text{r}_\text{host-disk} = \dmhostdisksubtractrest$. We then calculate the ratio of observables
\begin{equation*}
    \frac{\tau_{\mathrm{proper},\SI{1}{\giga\hertz}}}{3 (\mathrm{DM}^\text{r}_\mathrm{host-disk})^2} \lesssim \SI{4 +- 3e-7}{\milli\s\per\parsec\squared\centi\m\tothe{6}} \propto \widetilde{F}G.
\end{equation*}
This ratio characterizes the efficiency of the scattering along the line of sight. It is proportional to the product of the fluctuation parameter $\widetilde{F}$ and an order-unity geometric factor $G$. The proportionality constant is $\Gamma(7/6) r_e^2 c^3 \nu^{-4}$, where $\Gamma(7/6) \approx \num{0.9277}$, $c$ is the speed of light, $r_e = \SI{2.8}{\femto\meter}$ is the classical electron radius, and $\nu$ is the frequency at which the scattering is observed~\cite{2016arXiv160505890C}. This proportionality constant captures the microphysics and the frequency dependence of the scattering and relates it to the ratio of observables. The bulk properties of the gas are captured by $\widetilde{F}$, which depends on the volume filling factor of gas cloudlets, the size distribution of cloudlets doing the scattering, the size of the density variations within a cloudlet, and the inner/outer scales of the turbulence~\cite{2021ApJ...911..102O}. For the Milky Way's disk, typical values of $\widetilde{F}$ range from \SIrange{0.001}{1}{\parsec\tothe{-2/3} \kilo\meter\tothe{-1/3}} for low-latitude sightlines, roughly corresponding to scattering-$\text{DM}^2$ ratios of \SIrange[retain-unity-mantissa=false]{1e-8}{1e-5}{\milli\s\per\parsec\squared\centi\m\tothe{6}} \cite{2021ApJ...911..102O}. $G$ can vary by an order of magnitude because it depends on the relative position of the scattering media to the source and observer, which is poorly constrained for extragalactic sources of scattering. For example, for the geometry of a homogeneous scattering medium between the FRB and the edge of the host galaxy and a distant observer at infinity, $G = 1$. However, for a spiral arm of thickness $L \approx \SI{1}{\kilo\parsec}$ at a distance $d \approx \SI{10}{\kilo\parsec}$ in front of the FRB, $G = L/d \approx \num{0.1}$. In conclusion, the host DM and scattering upper limit are consistent with expectations from a host-galactic disk with Milky Way-like density fluctuations. These properties are suggestive of a source close to the host galaxy's plane as opposed to an FRB progenitor significantly displaced from the host galaxy's disk.

Another interpretation is that the DM excess is partially contributed by the source's local environment. The DM excess observed is not extreme: it is only a factor of two greater than the median measured in population studies ($\text{DM}_{\text{host}} \approx \SI{145}{\parsec\centi\meter\tothe{-3}}$~\cite{2022MNRAS.509.4775J}). Furthermore, the upper limit on the scattering timescale and low RM are not outliers within the diverse population of FRBs. In this scenario, the FRB could be produced by a progenitor significantly displaced from the host galactic plane relative to the electron scale height (\SI{1.57+-0.15}{\kilo\parsec}), reducing the host disk contribution to a fraction of our estimate (\dmhostestimate). This displacement would imply an old progenitor since young progenitors typically have low scale heights, \SI{\sim30}{\parsec} and \SI{100}{\parsec}, for young magnetars and massive stars respectively~\cite{2014ApJS..212....6O,1979ApJS...41..513M}).

\subsection*{Polarisation Analysis}
\label{sec:polarisation_analysis} 
The polarisation analysis follows a similar procedure to that previously applied to other FRBs detected by CHIME/FRB~\cite{fab+20,2021ApJ...910L..18B}. In particular, an initial RM estimate is made by applying RM-synthesis \cite{b66,bb05} to the Stokes $Q$ and $U$ data of the burst. This initial estimate is then further refined through a judicious selection of time and frequency limits that optimize the S/N of the polarised signal. We then apply a Stokes $QU$-fitting routine that directly fits for the modulation between Stokes $Q$ and $U$ from Faraday rotation but is further extended to capture additional features in the Stokes spectrum. 

We analyse FRB 20210603A using the CHIME/FRB polarisation pipeline, identical to that recently employed on FRB 20191219F \cite{Mckinven_2021}. We determine an $\text{RM}=\SI{-219.00 \pm 0.01 }{\radian\per\m\squared}$ and find the lower limit of the linear polarised fraction ($\Pi_\text{L}$) differs between the top ($\gtrsim$\SI{96}{\percent} at \SI{800}{\mega\hertz}) and the bottom of the CHIME band ($\gtrsim$\SI{87}{\percent} at \SI{400}{\mega\hertz}). This is counteracted by a very small but changing circular polarised fraction that becomes more significant at the bottom of the band. While this result may reflect the intrinsic properties of the burst at the source or be an imprint of some unknown propagation effect \cite{Vedantham2019,Gruzinov2019,Beniamini2022}, it is also not possible to rule out instrumental effects such as cross-polarisation between CHIME's orthogonal feeds. For this reason, we do not report on the circular polarisation and conservatively set our $\Pi_\text{L}$ measurements as lower bounds (see Table~\ref{table:frbparams}).

The Galactic $\text{RM}_{\text{MW}}=\SI{-22.4 \pm 0.3}{\radian\per\m\squared}$ contribution can be estimated from recent all-sky Faraday Sky maps \cite{Hutschenreuter2021}. 
The RM contribution of Earth's ionosphere, $\text{RM}_{\text{iono}}=+\SI{1.4}{\radian\per\m\squared}$, is determined from the \texttt{RMextract} package~\cite{Mevius2018b}. The uncertainty on this value is not provided, however, the variability in $\text{RM}_{\text{iono}}$ is expected to be $\lesssim +\SI{1}{\radian\per\m\squared}$ based on observations of pulsars and repeating FRB sources. 

Given that the Galactic pulsar population preferentially occupies the Milky Way disk, this similarity, while not ruling out alternative scenarios, is consistent with the notion that FRB 20210603A resides in or near the disk component of its host galaxy. Extended Data Fig.~\ref{fig:dm_rm_dist_scatt} further explores this analysis by locating our $\text{DM}_\text{host}$, $\envert{\text{RM}_\text{host}}$ and $\tau_{\text{scatt}}$ estimates of FRB 20210603A within the equivalent phase space of the Galactic pulsar sample. Galactic pulsar data are obtained from the latest Australia Telescope National Facility (ATNF) pulsar catalogue \cite{Manchester2005} using the \texttt{psrqpy} package~\cite{psrqpy}. FRB~20210603A occupies a well sampled region of this phase space, however, the distribution is also seen to be highly dependent on the Galactic latitude. We estimate a quasi-latitude value for FRB 20210603A, determined from a simple transformation of the inclination angle of the host galaxy (i.e., $\SI{4}{\degree} \leq \SI{90}{\degree}-\rm{inclination\, angle} \leq \SI{10}{\degree}$), and find that the average pulsar properties of $\text{DM}$, $\envert{\text{RM}}$ and $\tau_{\text{scatt}}$ at this equivalent latitude agree well with what is observed from FRB 20210603A. The agreement is further improved by rescaling $\text{DM}$, $\envert{\text{RM}}$ to account for the larger disk mass of the host galaxy relative to the Milky Way. This scaling factor corresponds to the ratio of the disk mass of the host galaxy and Milky Way and is found to be $\del{\text{M}_\text{host}^\star/\text{M}^\star_\text{MW}}^{1/3} = $\massscalingfactor ~(See Dispersion and Scattering analysis). Such a result suggests that most of the observed $\text{DM}_\text{host}$, $\envert{\text{RM}_\text{host}}$ and $\tau_\text{scatt}$ observed from FRB 20210603A can be supplied by the host galaxy ISM with little additional contribution needed from the source's local environment. 

\subsection*{Host Galaxy Analysis}
\label{sec:galaxy_analysis}

Optical images of SDSS J004105.82+211331.9 were taken with the CFHT MegaCam using the wide-band \textit{gri} filter. The data were reduced using the standard bias, dark, and flat corrections using the Elixir pipeline \cite{2004Magnier,2013Prunet}. Several exposures were combined using this filter to create an image with a total exposure of \SI{2500}{\s}. 

The half-light radius of the host galaxy was determined using the given Petrosian radii fluxes provided by SDSS Data Release 12 \cite{SDSS12} and Eq.~7 of \cite{PetroRadii}. The half-light radius in the $r$-band using these values was found to be \SI{8.2 +- 0.9}{\kilo\parsec}. Furthermore, the SDSS-provided apparent magnitude in the $r$-band was corrected for Milky Way extinction using the model from Fitzpatrick \& Massa 2007 \cite{2007FM}; this gave us an absolute magnitude of \num{-22.03 +- 0.02} after k-corrections \cite{2010MNRAS.405.1409C}.

In addition to imaging, we conducted Gemini spectroscopic observations consisting of two \SI{1000}{\s} exposures, one centred at \SI{6750}{\angstrom} and the other centred at \SI{6650}{\angstrom}. This wavelength offset was to account for the gap between the detectors. The images were reduced using standard bias and flat corrections, and combined using the Gemini \texttt{IRAF/PyRAF} package tools \cite{1986IRAF, 1993IRAF}. Using the same package, we also wavelength- and flux-calibrated the spectrum, and accounted for skylines and cosmic rays in the data. We extract spectra with various aperture sizes along the galaxy. The redshift was determined by extracting a spectrum from a \SI{1}{\arcsec} wide aperture centred at the central coordinates of the host galaxy. Due to the edge-on orientation of the galaxy, almost all of the galaxy's light falls within the slit, and the effect of slit corrections on the measured fluxes are negligible~(see Extended Data Fig.~\ref{fig:galaxy_spectra}).

The H$\alpha$ and the redwards line of the N~{\small II} doublet (rest wavelengths of \SI{6564.6}{\angstrom} and \SI{6585.2}{\angstrom}) are some of the most detectable lines (Extended Data Fig.~\ref{fig:galaxy_spectra}). Other prominent lines are from Na and Mg absorption (rest wavelengths of \SI{5895.6}{\angstrom} and \SI{5176.7}{\angstrom}).
Fitting a linear combination of Gaussian line profiles to the H$\alpha$ and N~{\small II} lines yields a redshift of $z = \redshift$. The uncertainty in the spectroscopic redshift is dominated by the statistical uncertainties in the measured spectrum, which are normalized such that the reduced-$\chi^2$ of the residuals is \num{1}.

To further characterize the galaxy, we combine our Gemini spectra with archival 2MASS~\cite{2006AJ....131.1163S} and WISE photometry~\cite{2010AJ....140.1868W}.
We use the spectral-energy distribution (SED) fitting code \texttt{Prospector} to determine the stellar mass, metallicity, and star formation history of the galaxy~\cite{2021ApJS..254...22J}. Our modelling and analysis of this host galaxy closely follows a similar effort for FRB 20181030A~\cite{Bhardwaj_2021}. However, because the galaxy is nearly edge-on, dust extinction in the host-galactic centre reddens the observed emission. Therefore, we first correct the spectrum for extinction (see Eqs.~10 and 13 of \cite{Shao_2007}) due to its inclination of $\SI{83 +- 3}{\degree}$~\cite{2020ApJ...902..145K}. 

Our best-fit model is overlaid on the spectral and photometric data in Extended Data Fig.~\ref{fig:sed}. It assumes a delay-$\uptau$ star formation history $\propto t \exp\del{-{t}/\uptau}$, where $\uptau$ is the characteristic decay time and $t$ is the time since the formation epoch of the galaxy. We set five free parameters: present-day stellar mass, metallicity, $\uptau$, $t$, and the diffuse dust V-band optical depth (referred to as ``dust2'' in \texttt{Prospector}), which accounts for the attenuation of old stellar light. We use $\uptau$ and $t$ as determined by \texttt{Prospector} to calculate the mass-weighted age of the galaxy. Additionally, we used a standard dust attenuation model~\cite{2000calzetti}, and enabled nebular emission and dust emission ~\cite{2017Byler,2007DraineLi}.

Before sampling the likelihood, we choose reasonable priors for each free parameter (Extended Data Table~\ref{table:priors}). We use Eq.~(6) of~\cite{2010Bernardi} to obtain an initial estimate of the galaxy's mass, and to set a weak prior on the mass range.
\begin{equation}
    \log_{10}(\text{M}_\text{host}^\star/\text{M}_{\odot}) = \num{1.097}(g-r) - \num{4.06} - \num{0.4}(M_r - \num{4.97}) - \num{0.19}z,
\end{equation}
where $g$ and $r$ are the apparent magnitudes in the $g$-band and $r$-band filters, $M_r$ is the absolute magnitude in the $r$-band, and $z$ is the redshift. The prior on $t$ was cut off at \SI{12}{\giga\year} because the age of the Universe at $z =\redshift$ is only \SI{\sim12}{\giga\year}. The prior on $\text{Z}/\text{Z}_\odot$ and $\uptau$ were set according to recommendations in \texttt{Prospector}~\cite{2021ApJS..254...22J}. Using these priors, we obtain the fit plotted in Extended Data Fig.~\ref{fig:sed} and list the results in Table~\ref{table:frbparams}. 

Finally, to determine the galaxy-integrated SFR, we extract a spectrum with an aperture of \SI{10}{\arcsec} in diameter, encompassing all of the galaxy's light within our half-light radius of $\SI{\sim2.5}{\arcsec}$. We calculate the total SFR of the host galaxy using the intensity and line width of the H$\alpha$ line~\cite{1994ApJ...435...22K}: 
\begin{equation}
    \text{SFR} = \num{7.9e-42} \del{\frac{\text{L}_{\text{H}\alpha}}{\si{\erg\per\s}}}\frac{\text{M}_\odot}{\si{\year}},
\end{equation}
where $\text{L}_{\text{H}\alpha}$ is the flux-derived luminosity of the $\text{H}\alpha$ emission from our Gemini data. To correct our luminosity measurement for extinction we apply the inclination-angle dependent correction as well as the inclination-independent correction, parameterized as dust2 in \texttt{Prospector}. The latter quantifies the amount of V-band extinction of old stellar light in the host galaxy. Optical reddening is characterized by using $\textrm{R}_\text{V}$ = $\textrm{A}_\text{V}$/E(B-V), where E(B-V) is the colour index of the galaxy and $\textrm{A}_\text{V}$ is the extinction in the V-band; this equation is thus the ratio of total to selective extinction in the V-band \cite{1999Fitz}. The dust extinction is taken to be $\textrm{A}_\text{V} = \num{1.086}\times \text{dust2}$~\cite{conroy2009a}, where we take dust2 to be the best-fit value of \num{0.79}. With an $\textrm{R}_\text{V}$ value of \num{3.1} \cite{1999Fitz}, we calculated E(B-V) to be \num{0.28}. The H$\alpha$ extinction coefficient can be calculated using $\textrm{A}_{\text{H}\alpha} = \textrm{R}_{\text{H}\alpha} \times$ E(B-V) where we take $\textrm{R}_{\text{H}\alpha} = \num{2.45}$; this value is within the range of values predicted by several different extinction models \cite{2011FS,2000calzetti,2007FM,o_donnell1994r}. The inclination-independent attenuation results in the H$\alpha$ flux being attenuated by a factor of $\exp\del{\textrm{A}_{\text{H}\alpha}} = \dustcorr$. Correcting the galaxy-integrated H$\alpha$ flux for extinction yielded a total SFR of $\num{0.24\pm0.06}\,\Msun \si{\per\year}$. 

\subsection*{Disk Chance Coincidence Probability}
While FRB 20210603A was ostensibly localized to the disk of its host galaxy, it is possible that the progenitor is actually a halo object (as in the case of the globular cluster host of FRB 20200120E~\cite{2022Natur.602..585K}) coincidentally aligned with the disk in projection. The probability that this occurs by a chance coincidence is small: we estimate this as the ratio of the solid angles subtended by the disk and halo, that is $P_{cc} \approx \Omega_\mathrm{disk}/\Omega_\mathrm{halo} \approx 10^{-3}$. The angular area of the nearly edge-on disk can be approximated as an ellipse with major and minor axes of 15 and \SI{2.7}{\arcsec} respectively, while the area of the halo can be approximated as a circle of radius $r_\mathrm{vir} \approx \text{M}_\text{host}^\star/\text{M}^\star_\text{MW} r_\mathrm{vir, MW} \approx \SI{\sim 280}{\kilo\parsec}$ estimated by scaling up the Milky Way's virial radius $r_\mathrm{vir, MW}\approx\SI{200}{\kilo\parsec}$~\cite{2006MNRAS.369.1688D}. This low chance coincidence probability of $10^{-3}$ implies a robust association with the disk and favours progenitor models involving disk populations over halo populations.

\backmatter


\bmhead{Data Availability}

Calibrated visibilities, dynamic spectra for producing figures, and MCMC chains for the localization analysis are available upon request and will be hosted by the time of publication as downloadable \texttt{HDF5} files on the CHIME/FRB CANFAR repository at \url{https://www.canfar.net/storage/list/AstroDataCitationDOI/CISTI.CANFAR/24.0086/data}. Optical images, spectra, and photometric data are immediately available as \texttt{fits} files at \url{https://github.com/tcassanelli/frb-vlbi-loc}.

\bmhead{Code Availability}

The code used for beamforming, VLBI localization, and polarisation analysis are available on Github:  \url{https://github.com/CHIMEFRB/baseband-analysis}. The scattering timescale has been measured using \texttt{fitburst} \cite{2024ApJS..271...49F}, available at \url{https://github.com/CHIMEFRB/fitburst}. Code for interpreting burst properties, and producing the figures and tables in this manuscript from the results of our analyses is available at \url{https://github.com/tcassanelli/frb-vlbi-loc}. In our analyses, we also make use of open-source software including
\texttt{astropy} \cite{2018AJ....156..123A}, \texttt{baseband} \cite{marten_van_kerkwijk_2020_4292543}, \texttt{difxcalc11} \cite{2016ivs..conf..187G}, \texttt{matplotlib} \cite{2007CSE.....9...90H}, \texttt{numpy} \cite{2020Natur.585..357H}, \texttt{scipy} \cite{2020SciPy-NMeth} , \texttt{h5py} \cite{hdf5}, \texttt{emcee} \cite{2013PASP..125..306F}, and \texttt{corner} \cite{Foreman-Mackey2016}, \texttt{cartopy} \cite{Cartopy}, \texttt{IRAF} \cite{1986IRAF, 1993IRAF}, and \texttt{prospector} \cite{2021ApJS..254...22J}.

\bmhead{Acknowledgements}
We would like to dedicate this work to our colleague Dr.\,Jing Luo, who passed away on February \nth{15}, 2022. Jing worked for several years on the commissioning and maintenance of the 10-m telescope at the Algonquin Radio Observatory. His expertise in radio pulsars, radio observations, and hardware were indispensable in accomplishing this scientific milestone.

We acknowledge that CHIME is located on the traditional, ancestral, and unceded territory of the Syilx/Okanagan people. We are grateful to the staff of the Dominion Radio Astrophysical Observatory, which is operated by the National Research Council of Canada.  CHIME is funded by a grant from the Canada Foundation for Innovation (CFI) 2012 Leading Edge Fund (Project 31170) and by contributions from the provinces of British Columbia, Qu\'{e}bec and Ontario. The CHIME/FRB Project is funded by a grant from the CFI 2015 Innovation Fund (Project 33213) and by contributions from the provinces of British Columbia and Qu\'{e}bec, and by the Dunlap Institute for Astronomy and Astrophysics at the University of Toronto. Additional support was provided by the Canadian Institute for Advanced Research (CIFAR), McGill University and the McGill Space Institute thanks to the Trottier Family Foundation, and the University of British Columbia. The Dunlap Institute is funded through an endowment established by the David Dunlap family and the University of Toronto.

We would like to thank the staff of Dominion Radio Astrophysical Observatory, operated by the National Research Council of Canada, for gracious hospitality, support, and expertise. The ARO 10-m telescope operated by the University of Toronto. TONE is located at Green Bank Observatory, the Green Bank Observatory facility, which is supported by the National Science Foundation, and is operated by Associated Universities, Inc.~under a cooperative agreement. We would like to thank the staff at Green Bank Observatory for logistical support during the construction and operations of TONE.

This work is based on observations obtained at the international Gemini Observatory, a program of NSF's NOIRLab, which is managed by the Association of Universities for Research in Astronomy (AURA) under a cooperative agreement with the National  IRAF was distributed by the National Optical Astronomy Observatory, which was managed by the Association of Universities for Research in Astronomy (AURA) under a cooperative agreement with the National Science Foundation. PyRAF is a product of the Space Telescope Science Institute, which is operated by AURA for NASA.Science Foundation. on behalf of the Gemini Observatory partnership: the National Science Foundation (United States), National Research Council (Canada), Agencia Nacional de Investigaci\'{o}n y Desarrollo (Chile), Ministerio de Ciencia, Tecnolog\'{i}a e Innovaci\'{o}n (Argentina), Minist\'{e}rio da Ci\^{e}ncia, Tecnologia, Inova\c{c}\~{o}es e Comunica\c{c}\~{o}es (Brazil), and Korea Astronomy and Space Science Institute (Republic of Korea).

This work is also based on observations obtained with MegaPrime/MegaCam, a joint project of CFHT and CEA/DAPNIA, at the Canada-France-Hawaii Telescope (CFHT) which is operated by the National Research Council (NRC) of Canada, the Institut National des Science de l'Univers of the Centre National de la Recherche Scientifique (CNRS) of France, and the University of Hawaii. The observations at the Canada-France-Hawaii Telescope were performed with care and respect from the summit of Maunakea which is a significant cultural and historic site. 

IRAF was distributed by the National Optical Astronomy Observatory, which was managed by the Association of Universities for Research in Astronomy (AURA) under a cooperative agreement with the National Science Foundation. PyRAF is a product of the Space Telescope Science Institute, which is operated by AURA for NASA.

\allacks

\bmhead{Author contributions}

T.C.\ led the full instrument adaptation, triggering system, and data management of the ARO10 telescope, wrote in the main text, methods sections, and prepared several figures and tables. C.L.\ wrote the VLBI software correlator and analysis pipeline used to localize the FRB, designed and built the digital backend of the TONE array, and led the data analysis, scientific interpretation, and writing of the manuscript. P.S.\ led the design, construction, commissioning, and data acquisition of all aspects of the TONE telescope, and contributed significantly to the scientific interpretation and writing of the manuscript. J. M. P.\ designed and installed the maser hardware, characterized the clock stabilization system, and made foundational contributions to the array calibration pipelines used at TONE and CHIME. S.C.\ led the calibration, reduction, and analysis of optical follow-up data. All authors from the CHIME/FRB collaboration played either leadership or significant supporting roles in one or more of: the management, development, construction, commissioning, and maintenance of CHIME, the CHIME/FRB instrument, the ARO10 instrument, the TONE instrument, their respective software data pipelines, and/or the data analysis and preparation of this manuscript.


\bmhead{Competing Interests}. The authors declare that they have no competing financial interests.



\newpage

\section*{Tables}
\FloatBarrier

\begin{table}
    \centering
    \caption{\textbf{Measured and Derived Parameters associated with FRB 20210603A and its host galaxy}. Properties derived from radio and optical follow-up data are listed in the top and bottom halves of the table respectively. Parameters which are derived from external models or measurements are indicated with daggers. ($z_\text{phot}$, DM, $\uptau$, and RM$_{\text{iono}}$ predictions~\cite{2002astro.ph..7156C, 2003astro.ph..1598C,Mevius2018b,2020ApJ...888..105Y,2021MNRAS.505.5356B,Hutschenreuter2021}).
    }
    \label{table:frbparams}
    \begin{tabularx}{\textwidth}{l|l}
        \hline
        \textbf{Parameter} & \textbf{Value} \\ 
        \hline
        Right ascension $\upalpha$ (ICRS) & \radecimal \\ 
        Declination $\updelta$ (ICRS) & \decdecimal \\ 
        CHIME arrival time at ($\SI{400}{\mega\hertz} $) & 2021-06-03 15:51:34.431652 UTC \\
        Dispersion measure (DM) & \dmfrb \\ 
        $\text{DM}_\text{MW-NE2001}^\dagger$ & \dmmwne \\ 
        $\text{DM}_\text{MW-halo}^\dagger$ & \dmmwhalo  \\ 
        $\text{DM}_\text{cosmic}$ & \dmcosmic \\
        $(\text{DM}_\text{host})/(1+z) = (\text{DM}_\text{host-disk} + \text{DM}_\text{host-halo})/(1+z) $ & \dmhostbudget \\
        $\text{RM}$ & \SI{-219.00 \pm 0.01 }{\radian\per\m\squared}  \\
        $\text{RM}_{\text{MW}}^\dagger$ & \SI{-22.4 \pm 0.3}{\radian\per\m\squared} \\ 
        $\text{RM}_{\text{iono}}^\dagger$ & +\SI{1.4}{\radian\per\m\squared}  \\ 
        $\Pi_\text{L-\SI{800}{\mega\hertz}}$ & $\gtrsim\SI{96}{\percent}$ \\
        $\Pi_\text{L-\SI{400}{\mega\hertz}}$ & $\gtrsim\SI{87}{\percent}$  \\
        $\tau_\text{\SI{600}{\mega\hertz}}$ & $\SI{165 \pm 3}{\micro\second}$ \\
        $\tau_\text{\SI{600}{\mega\hertz}-NE2001}^\dagger$ & \SI{1.02}{\micro\second} \\
        Fluence  & $\SI{64.4 +- 6.5}{\jansky \milli\second}$ \\
        Flux density  & $\SI{64.9 +- 6.5}{\jansky}$ \\
        Specific energy & $\SI{5.7e31}{\erg/\hertz}$ \\
        Specific luminosity & $\SI{5.8e34}{\erg\per\s\per\hertz}$ \\
        Band-averaged pulse FWHM& $\SI{740}{\micro\second}$ \\
        \hline
        Spectroscopic redshift, $z$ & \redshift \\
        Photometric redshift, $z_{\text{phot}}^{\dagger}$ & $\num{0.175 +- 0.0133}$ \\
        Inclination angle & $\SI{83 \pm 3}{\degree}$ \\
        Present-day stellar mass, $\log(\text{M}_\text{host}^{\star}/\text{M}_{\odot})$ &  $\num{10.93}^{+0.04}_{-0.04}$ \\
        Metallicity, $\log\del{\text{Z}/\text{Z}_{\odot}}$ & $\num{-0.22}^{+0.05}_{-0.04}$ \\
        Mass-weighted age & $\num{4.32}^{+0.73}_{-0.75}$ \si{\giga\year}  \\
        Total star formation rate (SFR) & $\gtrsim 0.24 \pm 0.06 M_{\odot}  \text{yr}^{-1}$ \\
        Projected offset & \SI{7.2}{\kilo\parsec} \\ 
        $r$-band half-light radius & \SI{8.2 +- 0.9}{\kilo\parsec} \\
        Absolute $r$-band magnitude  & \num{-22.03 +- 0.02} \\
        E(B-V) & \num{0.28} \\
        \hline
    \end{tabularx}
\end{table}

\begin{table}
    \centering
    \caption{\textbf{A summary of the properties of the CHIME, ARO10, and TONE stations}. System equivalent flux density (SEFD) at ARO10 was calculated with a set of Crab GPs \cite{2022AJ....163...65C}. The SEFD and FoV of TONE have been computed from a drift scan observation of Taurus-A \cite{tonesystem}. CHIME SEFD at Dec $\SI{+22}{\degree}$ has not been calculated, but its system temperature has been extensively studied in \cite{2022ApJS..261...29C}.}
    \label{tab:telescope_properties}
    \begin{tabular}{lccc}
        \hline
        \textbf{Property} & \textbf{CHIME} & \textbf{ARO10} & \textbf{TONE} \\ \hline
        SEFDs $S_{\mathrm{sys}}$ at Dec $\SI{+22}{\degree}$ & -- & \SI{\sim1.7}{\kilo\jansky}  & \SIrange{\sim20}{40}{\kilo\jansky} \\
        Field of view (at \SI{600}{\mega\hertz}) & \SI{\sim110}{\degree} N-S, \SI{1.74}{\degree} E-W &  \SI{3.59}{\degree} & \SIrange{\sim6}{11}{\degree}\\
        Processed frequency channels & \num{916}  & \num{1024} & \num{1024} \\
        Baseline length & -- & $b_\mathrm{CA}=\SI{3074}{\kilo\m}$ & $b_\mathrm{CT}=\SI{3332}{\kilo\m}$ \\
        Longitude (\si{\deg}) & \num{-119.6237} & \num{-78.0701} & \num{-79.8452} \\
        Latitude (\si{\deg}) & \num{49.3207} &  \num{45.9556} & \num{38.4293} \\
        \hline
    \end{tabular}
\end{table}

\renewcommand{\tablename}{Extended Data Table} 
\setcounter{table}{0}

\begin{table}
    \centering
    \caption{\textbf{Priors set for SED modeling.} The parameters here are used for modeling the host galaxy with a delayed-$\tau$ model as implemented in \texttt{Prospector}.}
    \label{table:priors}
    \begin{tabular}{lll}
        \hline
        \textbf{Parameter} & \textbf{} & \textbf{Prior $\sbr{\text{min}, \text{max}}$} \\ 
        \hline
        $\log(M^{\star}/M_{\odot})$ & Present-day Stellar Mass & Log Uniform $\sbr{\num{10}, \num{12}}$ \\
        $\log\del{Z/Z_{\odot}}$ & Metallicity & Top Hat $\sbr{-2, 0.19}$\\
        $t$ & Time since formation (Gyr) & Top Hat $[0.1, 12]$ \\
        $\uptau$ & Star formation characteristic decay rate (Gyr) & Log Uniform $\sbr{0.3, 15}$\\
        dust2 & Diffuse dust V-band optical depth & Top Hat $\sbr{0,3}$ \\ \hline
    \end{tabular}
\end{table}

\clearpage
\pagebreak

\section*{Figure Legends/Captions}

\FloatBarrier

\renewcommand{\figurename}{Extended Data Fig.} 
\setcounter{figure}{0}

\begin{figure}
    \centering
    \includegraphics[width=0.9\textwidth, keepaspectratio]{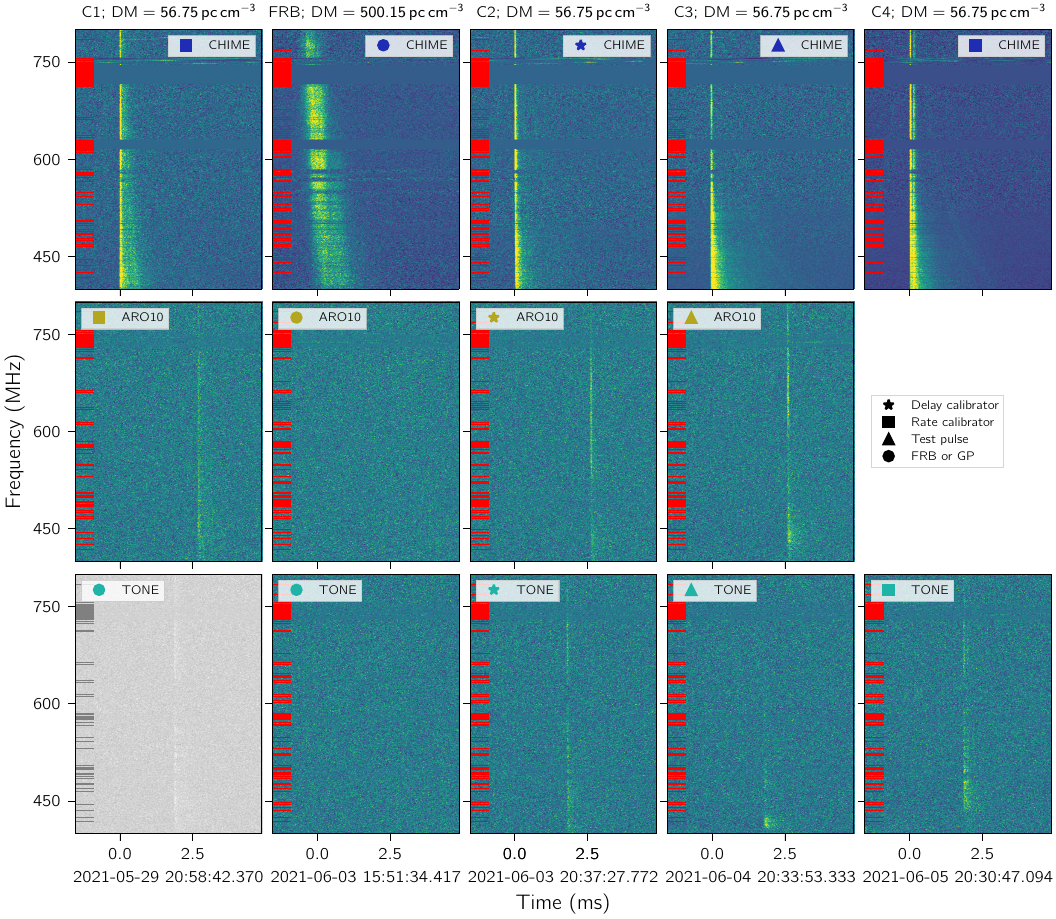}
    \caption{\textbf{Dynamic spectra of all observations.} At each VLBI station we recorded five single pulses (including the FRB): Crab GPs which we refer to as C1--C4 in the several days surrounding FRB 20210603A. Each row corresponds to a different VLBI station (CHIME at the Dominion Radio Astrophysical Observatory, ARO10 at the Algonquin Radio Observatory, and TONE at the Green Bank Observatory). 
    Timestamps show site-local clocks aligned to within \SI{2.56}{\micro\s} at a reference frequency of \SI{800.}{\mega\hertz}. Though the FRB is too faint to be detected at the testbeds alone, it is robustly detected in cross-correlation with CHIME at both stations. The intensity was adjusted by normalizing its standard deviation and setting the colour scale limits to the \num{1} and \num{99} percentile values of the data. Waterfall plots are shown downsampled to a frequency resolution of \SI{390.625}{\kilo\hertz} and a time resolution of \SI{25.6}{\micro\s}. 
    The noisy radio frequency interference (RFI) channels in $\SIrange{700}{750}{\mega\hertz}$ correspond to the cellular communications bands and the RFI channels at $\approx\SI{600}{\mega\hertz}$ frequencies correspond to television transmission bands. These RFI channels were removed in our analysis and are highlighted with red strikes to the left of each waterfall plot.
    Symbols next to the telescope label in each waterfall plot indicate what each Crab pulse was used for. We use C2 on all baselines as a phase/delay calibrator, and C1 and C4 as rate calibrators for the CHIME-ARO10 and CHIME-TONE baselines respectively. We localized C3 as an end-to-end cross-check of our calibration solutions.}
    \label{fig:i_stokes_events}
\end{figure}

\begin{figure}
    \centering
    \includegraphics[width=0.75\textwidth]{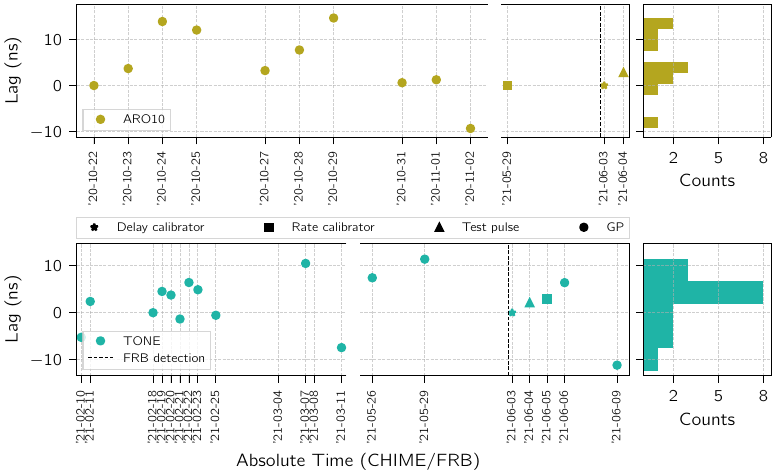}
    \caption{\textbf{Delay residuals measured from the CHIME-ARO10 and CHIME-TONE baselines.} The graph shows the empirical uncertainty obtained by analysing earlier data sets \protect\cite{2022AJ....163...65C,tonesystem}, with CHIME-ARO10 data shown in the top row and CHIME-TONE data showed in the bottom row. Each point corresponds to the residual delay after applying delay and phase corrections (CHIME-ARO10 is calibrated to 2020-10-22, and TONE is calibrated to 2021-02-18). The extracted delays have all been compensated for clock errors and for a clock rate error on the CHIME-ARO10 baseline.}
    \label{fig:empirical_error}
\end{figure}

\begin{figure}
    \centering
    \includegraphics[width=.4\textwidth, keepaspectratio]{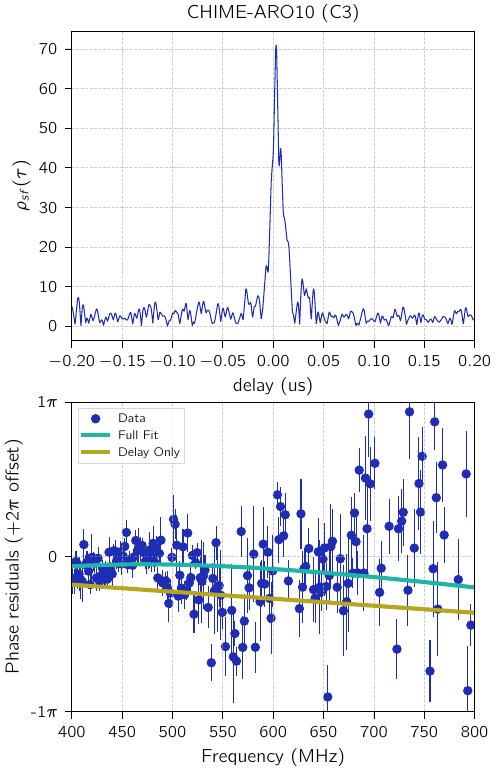}
    \hspace{.5cm}
    \includegraphics[width=.4\textwidth, keepaspectratio]{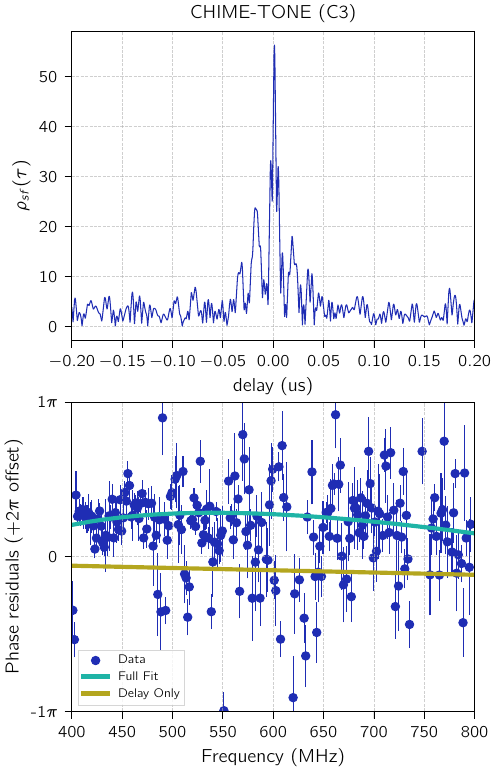}
    \caption{\textbf{Calibrated visibilities from the Crab pulsar giant pulse (C3) used to validate our calibration solutions.} We plot visibilities from the CHIME-ARO10 (left) and CHIME-TONE (right) baselines respectively. In each top panel, we plot the absolute value of the Fourier transform of the visibilities (i.e. the time-lag cross-correlation function $\rho\del{\tau}$ as a function of the delay referenced to the correlator pointing center). This shows a detection S/N exceeding \num{50} on each baseline. In each bottom panel we plot the phase of the calibrated visibilities $\mathcal{V}\sbr{i,k}$, binned to \SI{1.6}{\mega\hertz} resolution, with $1\sigma$ phase errors estimated from off-pulse scans ($N=10$) plotted as $\sigma\sbr{i,k} / \mathcal{V}\sbr{i,k}$ (blue points). In the bottom panels we overlay the phase model (Eq.~\ref{eq:phase_model}) evaluated at the parameters which maximize $\mathcal{L}_\varphi$, where we have fit for the ionosphere and the positions simultaneously (green ``full fit'' curve), as well as the phase model evaluated at the $\mathcal{L}_\tau$ position at zero ionosphere (yellow ``delay only'' curve). Since our correlator pointing is the $\mathcal{L}_\tau$ position, we would then expect the yellow ``delay only'' curve to be flat; note that our plotting code automatically unwraps all of the phases in each bottom panel by some amount automatically chosen to reduce phase wrapping, explaining the very small deviation from zero delay.}
    \label{fig:visibilities_crab}
\end{figure}

\begin{figure}
    \centering
    \includegraphics[width=.75\textwidth]{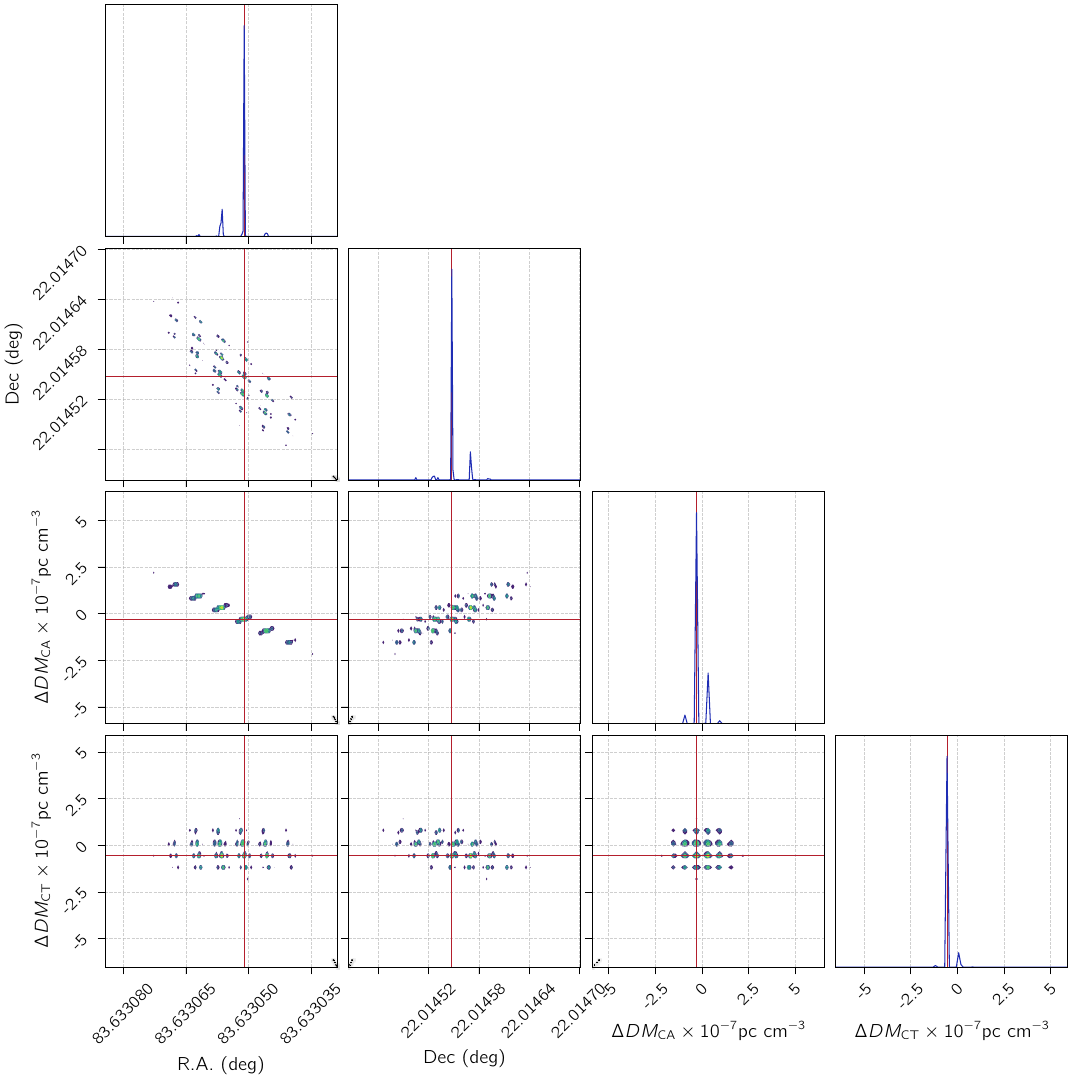}
    \caption{\textbf{The localization posterior of the Crab pulse (C3) as a function of RA, Dec, and $\Delta \mathrm{DM_{CA}}$, and $\Delta \mathrm{DM_{CT}}$.} Due to the extremely sparse sampling of the $uv$-plane, we bypass traditional methods of VLBI imaging, and directly fit the visibilities $\mathcal{V}\sbr{i,k}$. Owing to our wide bandwidth, we see that the ionosphere parameters $\Delta\mathrm{DM}$ are well-constrained even in the absence of external information (e.g., TEC maps or ionosphere priors). In the same spirit as a MCMC corner plot, each 2D plot shows the posterior marginalized over all except two axes. Calling these projections $P$, we colour evenly-spaced contours between $\log P = 0$ (the maximum value of each $P$ is normalized to \num{1}) and $\log P = -16$.}
    \label{fig:mcmc_corner_c3}
\end{figure}

\begin{figure}
    \centering
    \includegraphics{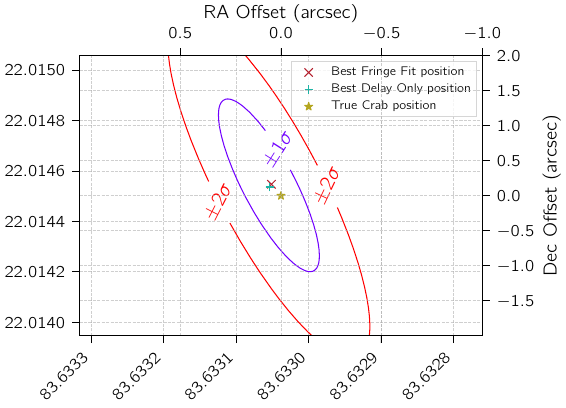}
    \caption{\textbf{Localization of C3 as an independent, end-to-end cross check of our VLBI calibration solution used to localize the FRB.} Due to the extremely sparse sampling of the $uv$-plane, we avoid traditional imaging. We compare two localization methods: a delay-space $\chi^2$-minimization of the residual delays left after calibration ($+$), and a visibility-space fitting of the phases ($\times$). Both methods agree to within the true position of the Crab (star) within systematic uncertainties (ellipses).}
    \label{fig:c3_loc}
\end{figure}

\begin{figure}
    \centering
    \includegraphics[width=.4\textwidth, keepaspectratio]{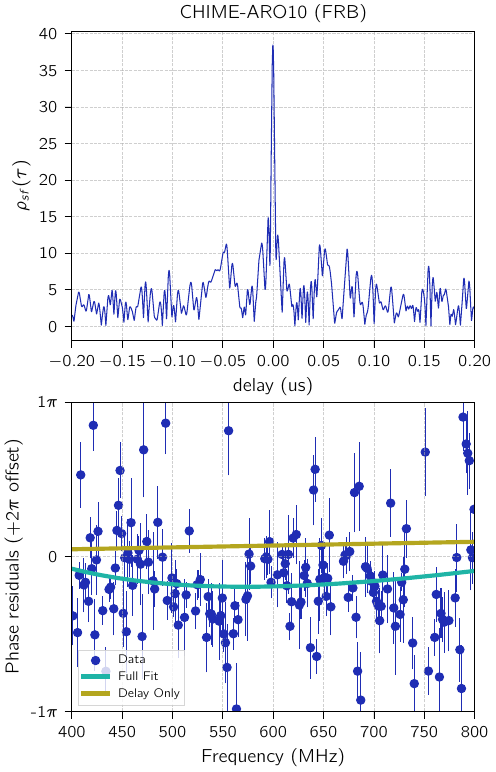}
    \hspace{.5cm}
    \includegraphics[width=.4\textwidth, keepaspectratio]{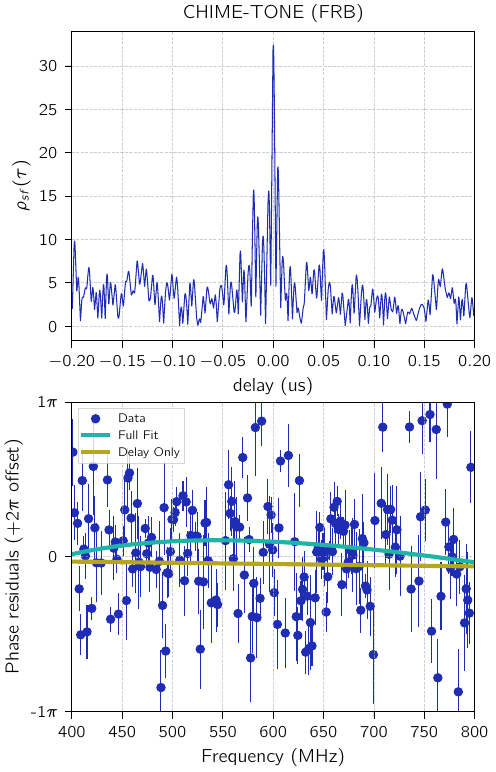}
    \caption{\textbf{Calibrated VLBI fringes on FRB 20210603A from the CHIME-ARO10 and CHIME-TONE baselines respectively.} We plot visibilities from the CHIME-ARO10 (left) and CHIME-TONE (right) baselines respectively. In each top panel, we plot the absolute value of the Fourier transform of the visibilities (i.e. the time-lag cross-correlation function $\rho\del{\tau}$ as a function of the delay referenced to the correlator pointing center). This shows a detection S/N exceeding \num{50} on each baseline. In each bottom panel we plot the phase of the calibrated visibilities $\mathcal{V}\sbr{i,k}$, binned to \SI{1.6}{\mega\hertz} resolution, with $1\sigma$ phase errors estimated from off-pulse scans ($N=10$) plotted as $\sigma\sbr{i,k} / \mathcal{V}\sbr{i,k}$ (blue points). In the bottom panels we overlay the phase model (Eq.~\ref{eq:phase_model}) evaluated at the parameters which maximize $\mathcal{L}_\varphi$, where we have fit for the ionosphere and the positions simultaneously (green ``full fit'' curve), as well as the phase model evaluated at the $\mathcal{L}_\tau$ position at zero ionosphere (yellow ``delay only'' curve). Since our correlator pointing is the $\mathcal{L}_\tau$ position, we would then expect the yellow ``delay only'' curve to be flat; note that our plotting code automatically unwraps all of the phases in each bottom panel by some amount automatically chosen to reduce phase wrapping, explaining the very small deviation from zero delay.}
    \label{fig:visibilities_frb}
\end{figure}

\begin{figure}
    \centering
    \includegraphics[width=.75\textwidth]{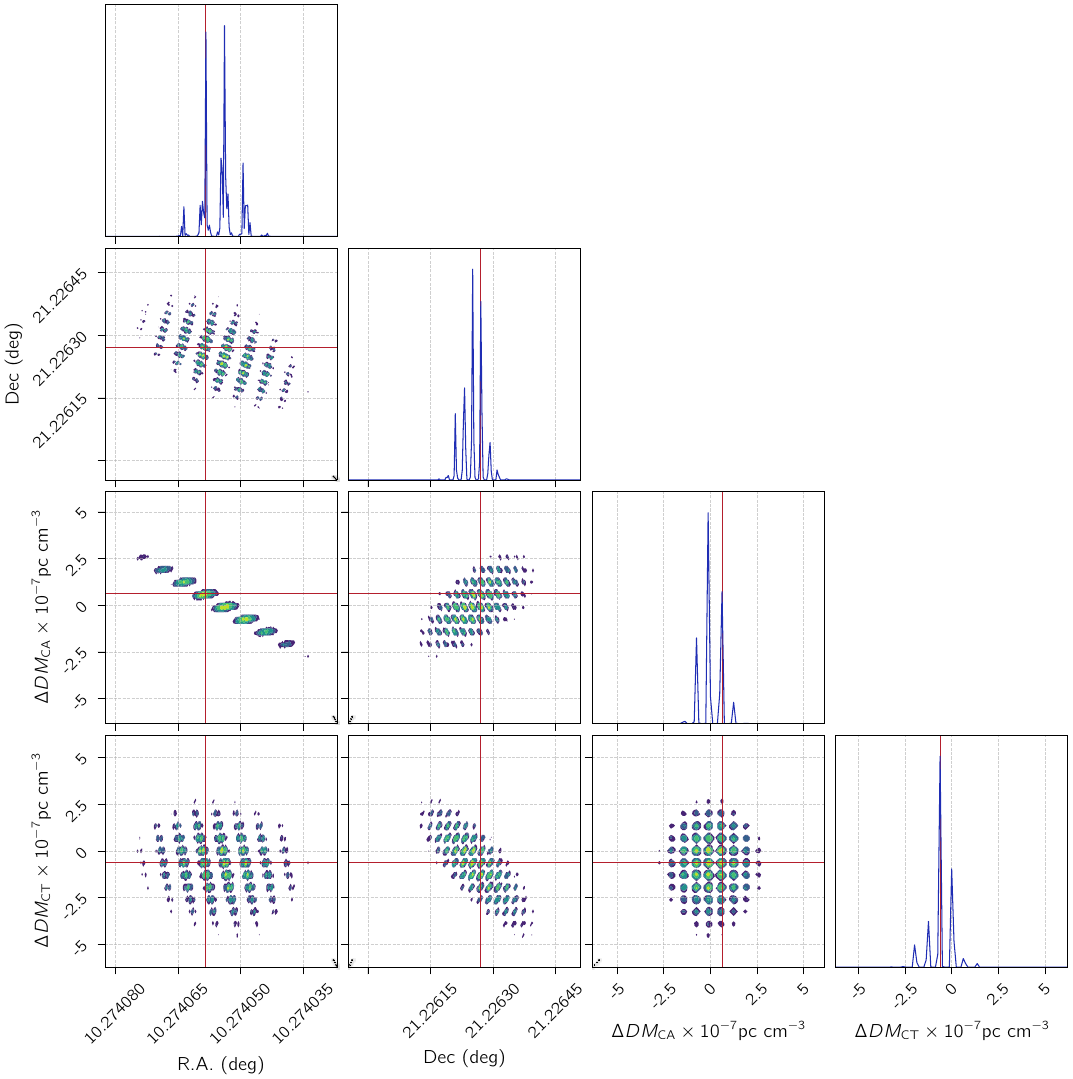}
    \caption{\textbf{The posterior localization contour of FRB 20210603A as a function of RA, Dec, and $\Delta \mathrm{DM_{CA}}$, and $\Delta \mathrm{DM_{CT}}$.} The ionosphere parameters $\Delta\mathrm{DM}$ are well-constrained even in the absence of external information (e.g., TEC maps or ionosphere priors). In the same spirit as a MCMC corner plot, each 2D plot shows the posterior marginalized over all except two axes. Calling these projections $P$, we colour evenly-spaced contours between $\log P = \num{0}$ (the maximum value of each $P$ is normalized to \num{1}) and $\log P = \num{-16}$.}
    \label{fig:mcmc_corner_frb}
\end{figure}

\begin{figure}
    \centering
    \includegraphics{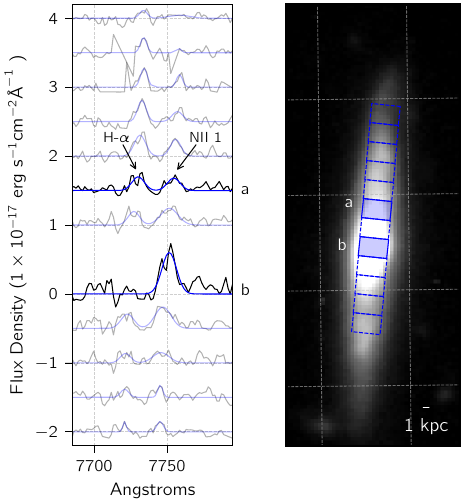}
    \caption{\textbf{Spatially resolved spectroscopy of the host galaxy.} Optical image and spatially-resolved spectra of the host galaxy of FRB 20210603A acquired using CFHT MegaCAM and Gemini long-slit spectroscopy respectively. Pixel intensities are scaled linearly and normalized to reduce the saturation evident in Figure~\ref{fig:galaxy_image}. All spectra are given offsets in increments of \SI[retain-unity-mantissa=false]{1e-17}{\erg\per\s\per\centi\m\squared\per\angstrom}. 
    One spectrum is extracted from the bulge of the galaxy (spectrum b, centered at 0). There are additional eleven spectra extracted from the FRB side of the galaxy (shown as positive offsets), and from the opposite side of the galaxy (shown as negative offsets), with offsets from the center of the galaxy in increments of $\SI{1}{\arcsec}$. 
    All spectra are extracted using an aperture size of $\SI{1.5}{\arcsec}\times\SI{1}{\arcsec}$, as represented on the galaxy image. Spectrum a is extracted within the vicinity of the FRB and represented by the shaded box a in the galaxy image. The twelve spectra and Gaussian fits to the H$\alpha$ and one of the N{\small II} emission lines, are plotted here after correcting for Milky-Way extinction.
    }
    \label{fig:galaxy_spectra}
\end{figure}

\begin{figure}
    \includegraphics[width=\textwidth,keepaspectratio]{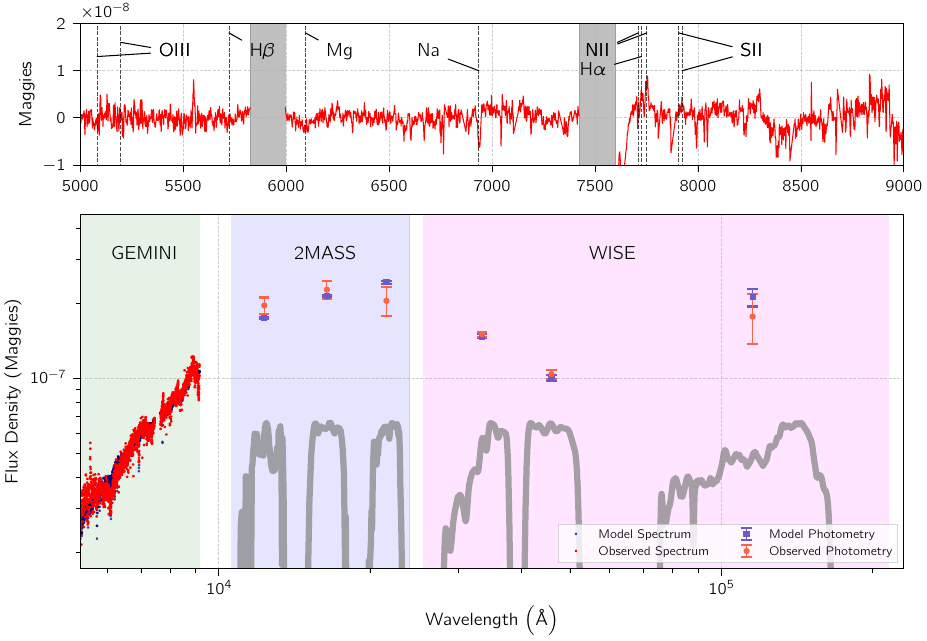}
    \caption{\textbf{Spectral energy distribution of host galaxy:} Gemini long-slit spectrum, integrated over the galaxy, with archival infrared photometry from 2MASS and WISE, plotted after correcting for extinction due to the host galaxy's inclination angle. Plotted alongside the observations (red) are the best-fit model (blue) from \texttt{Prospector}, and the relative passbands for the 2MASS J, H, and K$_s$ and WISE W1-W3 filters. Flux uncertainties are plotted by converting $1\sigma$ photometric errors reported by each catalogue.}
    \label{fig:sed}
\end{figure}

\begin{figure}
    \centering
    \includegraphics[width=0.75\textwidth]{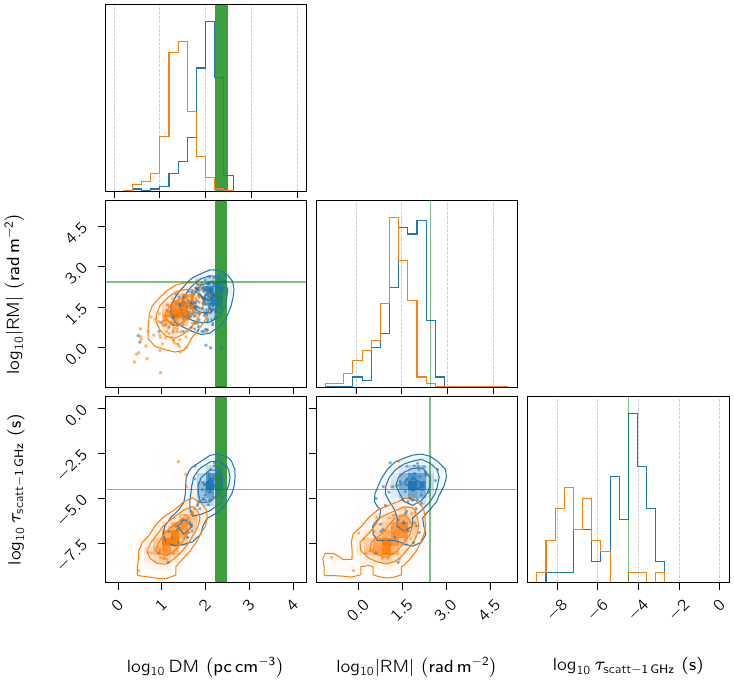}
    \caption{\textbf{A visualization of propagation effects due to the Milky Way's disk, as measured via the ATNF Pulsar Catalogue.} We plot joint distributions of $\text{DM}$, $\envert{\text{RM}}$ and $\tau_\text{scatt}$ for Galactic pulsars for two different latitude ranges: $\SI{4}{\degree} \leq \envert{b} \leq \SI{10}{\degree}$ (blue) and $\envert{b}\geq \SI{20}{\degree}$ (orange) taken from the ATNF Pulsar Catalogue~\cite{Manchester2005}. Contour lines indicate \num{1}, \num{2} and \num{3}$\sigma$ regions of this parameter space. Green regions/lines indicate estimates of equivalent quantities determined for the host galaxy of FRB 20210603A, namely: $\text{DM}_\text{host}$, $\envert{\text{RM}_\text{host}}$ and our upper limit on $\tau_\text{scatt}$. $\text{DM}_\text{host}$, $\envert{\text{RM}_\text{host}}$ and $\tau_\text{scatt}$ estimates are in the source frame with $\tau_\text{scatt}$ referenced at $\SI{1}{\giga\hertz}$ assuming a $\tau_\text{scatt}\propto \nu^{-4.4}$ relation used by ATNF. This shows that the burst properties of FRB 20210603A ($\text{DM}_{\text{host}}$, $\envert{\text{RM}_{\text{host}}}$ and $\tau_\text{scatt-\SI{1}{\giga\hertz}}$), once corrected for extragalactic contributions, are similar to that of low-latitude ($\SI{4}{\degree} \leq \envert{b} \leq \SI{10}{\degree}$) Galactic pulsars. 
    }
    \label{fig:dm_rm_dist_scatt}
\end{figure}

\clearpage
\pagebreak

\bibliography{frb-vlbi-loc-ref}

\begin{thebibliography}{100}
\expandafter\ifx\csname url\endcsname\relax
  \def\url#1{\burl{#1}}\fi
\expandafter\ifx\csname urlprefix\endcsname\relax\def\urlprefix{URL }\fi
\providecommand{\bibinfo}[2]{#2}
\providecommand{\eprint}[2][]{\url{#2}}
\providecommand{\doi}[1]{\url{https://doi.org/#1}}
\bibcommenthead

\bibitem{2018ApJ...863...48C}
\bibinfo{author}{{CHIME/FRB Collaboration}} \emph{et~al.}
\newblock \bibinfo{title}{{The CHIME Fast Radio Burst Project: System Overview}}.
\newblock \emph{\bibinfo{journal}{\apj}} \textbf{\bibinfo{volume}{863}}, \bibinfo{pages}{48} (\bibinfo{year}{2018}).

\bibitem{2022ApJS..261...29C}
\bibinfo{author}{{CHIME Collaboration}} \emph{et~al.}
\newblock \bibinfo{title}{{An Overview of CHIME, the Canadian Hydrogen Intensity Mapping Experiment}}.
\newblock \emph{\bibinfo{journal}{\apjs}} \textbf{\bibinfo{volume}{261}}, \bibinfo{pages}{29} (\bibinfo{year}{2022}).

\bibitem{2022AJ....163...65C}
\bibinfo{author}{{Cassanelli}, T.} \emph{et~al.}
\newblock \bibinfo{title}{{Localizing FRBs through VLBI with the Algonquin Radio Observatory 10 m Telescope}}.
\newblock \emph{\bibinfo{journal}{\aj}} \textbf{\bibinfo{volume}{163}}, \bibinfo{pages}{65} (\bibinfo{year}{2022}).

\bibitem{tonesystem}
\bibinfo{author}{{Sanghavi}, P.} \emph{et~al.}
\newblock \bibinfo{title}{{TONE: A CHIME/FRB Outrigger Pathfinder for localizations of Fast Radio Bursts using Very Long Baseline Interferometry}}.
\newblock \emph{\bibinfo{journal}{arXiv e-prints}} \bibinfo{pages}{arXiv:2304.10534} (\bibinfo{year}{2023}).

\bibitem{2021ApJ...910..147M}
\bibinfo{author}{{Michilli}, D.} \emph{et~al.}
\newblock \bibinfo{title}{{An Analysis Pipeline for CHIME/FRB Full-array Baseband Data}}.
\newblock \emph{\bibinfo{journal}{\apj}} \textbf{\bibinfo{volume}{910}}, \bibinfo{pages}{147} (\bibinfo{year}{2021}).

\bibitem{2024arXiv240305631L}
\bibinfo{author}{{Leung}, C.} \emph{et~al.}
\newblock \bibinfo{title}{{A VLBI Software Correlator for Fast Radio Transients}}.
\newblock \emph{\bibinfo{journal}{arXiv e-prints}} \bibinfo{pages}{arXiv:2403.05631} (\bibinfo{year}{2024}).

\bibitem{eubanks_proceedings_1991}
\bibinfo{author}{Eubanks, M.} \emph{et~al.}
\newblock \emph{\bibinfo{title}{Proceedings of the {U}.{S}. {Naval} {Observatory} {Workshop} on {Relativistic} {Models} for use in {Space} {Geodesy}}}  (\bibinfo{publisher}{U.S. Naval Observatory}, \bibinfo{year}{1991}).

\bibitem{SDSS12}
\bibinfo{author}{{Ahn}, C.~P.} \emph{et~al.}
\newblock \bibinfo{title}{{The Tenth Data Release of the Sloan Digital Sky Survey: First Spectroscopic Data from the SDSS-III Apache Point Observatory Galactic Evolution Experiment}}.
\newblock \emph{\bibinfo{journal}{\apjs}} \textbf{\bibinfo{volume}{211}}, \bibinfo{pages}{17} (\bibinfo{year}{2014}).

\bibitem{1998SPIE.3355..614B}
\bibinfo{author}{{Boulade}, O.} \emph{et~al.}
\newblock \bibinfo{editor}{{D'Odorico}, S.} (ed.) \emph{\bibinfo{title}{{Megacam: the next-generation wide-field imaging camera for CFHT}}}.
\newblock (ed.\bibinfo{editor}{{D'Odorico}, S.}) \emph{\bibinfo{booktitle}{Optical Astronomical Instrumentation}}, Vol. \bibinfo{volume}{3355} of \emph{\bibinfo{series}{Society of Photo-Optical Instrumentation Engineers (SPIE) Conference Series}}, \bibinfo{pages}{614--625} (\bibinfo{year}{1998}).

\bibitem{2020ApJ...902..145K}
\bibinfo{author}{{Kourkchi}, E.} \emph{et~al.}
\newblock \bibinfo{title}{{Cosmicflows-4: The Catalog of {\ensuremath{\sim}}10,000 Tully-Fisher Distances}}.
\newblock \emph{\bibinfo{journal}{\apj}} \textbf{\bibinfo{volume}{902}}, \bibinfo{pages}{145} (\bibinfo{year}{2020}).

\bibitem{2004PASP..116..425H}
\bibinfo{author}{{Hook}, I.~M.} \emph{et~al.}
\newblock \bibinfo{title}{{The Gemini-North Multi-Object Spectrograph: Performance in Imaging, Long-Slit, and Multi-Object Spectroscopic Modes}}.
\newblock \emph{\bibinfo{journal}{\pasp}} \textbf{\bibinfo{volume}{116}}, \bibinfo{pages}{425--440} (\bibinfo{year}{2004}).

\bibitem{2020A&A...641A...6P}
\bibinfo{author}{{Planck Collaboration}} \emph{et~al.}
\newblock \bibinfo{title}{{Planck 2018 results. VI. Cosmological parameters}}.
\newblock \emph{\bibinfo{journal}{\aap}} \textbf{\bibinfo{volume}{641}}, \bibinfo{pages}{A6} (\bibinfo{year}{2020}).

\bibitem{2022AJ....163...69B}
\bibinfo{author}{{Bhandari}, S.} \emph{et~al.}
\newblock \bibinfo{title}{{Characterizing the Fast Radio Burst Host Galaxy Population and its Connection to Transients in the Local and Extragalactic Universe}}.
\newblock \emph{\bibinfo{journal}{\aj}} \textbf{\bibinfo{volume}{163}}, \bibinfo{pages}{69} (\bibinfo{year}{2022}).

\bibitem{2006AJ....131.1163S}
\bibinfo{author}{{Skrutskie}, M.~F.} \emph{et~al.}
\newblock \bibinfo{title}{{The Two Micron All Sky Survey (2MASS)}}.
\newblock \emph{\bibinfo{journal}{\aj}} \textbf{\bibinfo{volume}{131}}, \bibinfo{pages}{1163--1183} (\bibinfo{year}{2006}).

\bibitem{2010AJ....140.1868W}
\bibinfo{author}{{Wright}, E.~L.} \emph{et~al.}
\newblock \bibinfo{title}{{The Wide-field Infrared Survey Explorer (WISE): Mission Description and Initial On-orbit Performance}}.
\newblock \emph{\bibinfo{journal}{\aj}} \textbf{\bibinfo{volume}{140}}, \bibinfo{pages}{1868--1881} (\bibinfo{year}{2010}).

\bibitem{2021ApJS..254...22J}
\bibinfo{author}{{Johnson}, B.~D.}, \bibinfo{author}{{Leja}, J.}, \bibinfo{author}{{Conroy}, C.} \& \bibinfo{author}{{Speagle}, J.~S.}
\newblock \bibinfo{title}{{Stellar Population Inference with Prospector}}.
\newblock \emph{\bibinfo{journal}{\apjs}} \textbf{\bibinfo{volume}{254}}, \bibinfo{pages}{22} (\bibinfo{year}{2021}).

\bibitem{2013PASP..125..306F}
\bibinfo{author}{{Foreman-Mackey}, D.}, \bibinfo{author}{{Hogg}, D.~W.}, \bibinfo{author}{{Lang}, D.} \& \bibinfo{author}{{Goodman}, J.}
\newblock \bibinfo{title}{{emcee: The MCMC Hammer}}.
\newblock \emph{\bibinfo{journal}{\pasp}} \textbf{\bibinfo{volume}{125}}, \bibinfo{pages}{306} (\bibinfo{year}{2013}).

\bibitem{2022Natur.606..873N}
\bibinfo{author}{{Niu}, C.~H.} \emph{et~al.}
\newblock \bibinfo{title}{{A repeating fast radio burst associated with a persistent radio source}}.
\newblock \emph{\bibinfo{journal}{\nat}} \textbf{\bibinfo{volume}{606}}, \bibinfo{pages}{873--877} (\bibinfo{year}{2022}).

\bibitem{2021ApJS..257...59C}
\bibinfo{author}{{CHIME/FRB Collaboration}} \emph{et~al.}
\newblock \bibinfo{title}{{The First CHIME/FRB Fast Radio Burst Catalog}}.
\newblock \emph{\bibinfo{journal}{\apjs}} \textbf{\bibinfo{volume}{257}}, \bibinfo{pages}{59} (\bibinfo{year}{2021}).

\bibitem{2002astro.ph..7156C}
\bibinfo{author}{{Cordes}, J.~M.} \& \bibinfo{author}{{Lazio}, T.~J.~W.}
\newblock \bibinfo{title}{{NE2001.I. A New Model for the Galactic Distribution of Free Electrons and its Fluctuations}}.
\newblock \emph{\bibinfo{journal}{arXiv e-prints}} \bibinfo{pages}{astro--ph/0207156} (\bibinfo{year}{2002}).

\bibitem{2003astro.ph..1598C}
\bibinfo{author}{{Cordes}, J.~M.} \& \bibinfo{author}{{Lazio}, T.~J.~W.}
\newblock \bibinfo{title}{{NE2001. II. Using Radio Propagation Data to Construct a Model for the Galactic Distribution of Free Electrons}}.
\newblock \emph{\bibinfo{journal}{arXiv e-prints}} \bibinfo{pages}{astro--ph/0301598} (\bibinfo{year}{2003}).

\bibitem{masui2015dense}
\bibinfo{author}{Masui, K.} \emph{et~al.}
\newblock \bibinfo{title}{Dense magnetized plasma associated with a fast radio burst}.
\newblock \emph{\bibinfo{journal}{Nature}} \textbf{\bibinfo{volume}{528}}, \bibinfo{pages}{523–525} (\bibinfo{year}{2015}).
\newblock \urlprefix\url{http://dx.doi.org/10.1038/nature15769}.

\bibitem{Akahori2016}
\bibinfo{author}{{Akahori}, T.}, \bibinfo{author}{{Ryu}, D.} \& \bibinfo{author}{{Gaensler}, B.~M.}
\newblock \bibinfo{title}{{Fast Radio Bursts as Probes of Magnetic Fields in the Intergalactic Medium}}.
\newblock \emph{\bibinfo{journal}{\apj}} \textbf{\bibinfo{volume}{824}}, \bibinfo{pages}{105} (\bibinfo{year}{2016}).

\bibitem{2021ApJ...908L..12T}
\bibinfo{author}{{Tendulkar}, S.~P.} \emph{et~al.}
\newblock \bibinfo{title}{{The 60 pc Environment of FRB 20180916B}}.
\newblock \emph{\bibinfo{journal}{\apjl}} \textbf{\bibinfo{volume}{908}}, \bibinfo{pages}{L12} (\bibinfo{year}{2021}).

\bibitem{2021AJ....161...81L}
\bibinfo{author}{{Leung}, C.} \emph{et~al.}
\newblock \bibinfo{title}{{A Synoptic VLBI Technique for Localizing Nonrepeating Fast Radio Bursts with CHIME/FRB}}.
\newblock \emph{\bibinfo{journal}{\aj}} \textbf{\bibinfo{volume}{161}}, \bibinfo{pages}{81} (\bibinfo{year}{2021}).

\bibitem{2022AJ....163...48M}
\bibinfo{author}{{Mena-Parra}, J.} \emph{et~al.}
\newblock \bibinfo{title}{{A Clock Stabilization System for CHIME/FRB Outriggers}}.
\newblock \emph{\bibinfo{journal}{\aj}} \textbf{\bibinfo{volume}{163}}, \bibinfo{pages}{48} (\bibinfo{year}{2022}).

\bibitem{2016JAI.....541005B}
\bibinfo{author}{{Bandura}, K.} \emph{et~al.}
\newblock \bibinfo{title}{{ICE: A Scalable, Low-Cost FPGA-Based Telescope Signal Processing and Networking System}}.
\newblock \emph{\bibinfo{journal}{Journal of Astronomical Instrumentation}} \textbf{\bibinfo{volume}{5}}, \bibinfo{pages}{1641005} (\bibinfo{year}{2016}).

\bibitem{2020JAI.....950014D}
\bibinfo{author}{{Denman}, N.} \emph{et~al.}
\newblock \bibinfo{title}{{A GPU Spatial Processing System for CHIME}}.
\newblock \emph{\bibinfo{journal}{Journal of Astronomical Instrumentation}} \textbf{\bibinfo{volume}{9}}, \bibinfo{pages}{2050014--2} (\bibinfo{year}{2020}).

\bibitem{2021zndo...5842660R}
\bibinfo{author}{{Renard}, A.} \emph{et~al.}
\newblock \bibinfo{title}{{Kotekan: A framework for high-performance radiometric data pipelines}} (\bibinfo{year}{2021}).

\bibitem{2017ursi.confE...4N}
\bibinfo{author}{Ng, C.} \emph{et~al.}
\newblock \bibinfo{editor}{None} (ed.) \emph{\bibinfo{title}{Chime/frb: An application of fft beamforming for a radio telescope}}.
\newblock (ed.\bibinfo{editor}{None}) \emph{\bibinfo{booktitle}{2017 XXXIInd General Assembly and Scientific Symposium of the International Union of Radio Science (URSI GASS)}}, \bibinfo{pages}{1--4} (\bibinfo{year}{2017}).

\bibitem{2015MNRAS.446..857L}
\bibinfo{author}{{Lyne}, A.~G.} \emph{et~al.}
\newblock \bibinfo{title}{{45 years of rotation of the Crab pulsar}}.
\newblock \emph{\bibinfo{journal}{\mnras}} \textbf{\bibinfo{volume}{446}}, \bibinfo{pages}{857--864} (\bibinfo{year}{2015}).

\bibitem{2017arXiv170808521D}
\bibinfo{author}{Deng, M.} \& \bibinfo{author}{Campbell-Wilson, D.}
\newblock \bibinfo{editor}{None} (ed.) \emph{\bibinfo{title}{The cloverleaf antenna: A compact wide-bandwidth dual-polarization feed for chime}}.
\newblock (ed.\bibinfo{editor}{None}) \emph{\bibinfo{booktitle}{2014 16th International Symposium on Antenna Technology and Applied Electromagnetics (ANTEM)}}, \bibinfo{pages}{1--2} (\bibinfo{year}{2014}).

\bibitem{2022JATIS...8a1019C}
\bibinfo{author}{{Crichton}, D.} \emph{et~al.}
\newblock \bibinfo{title}{{Hydrogen Intensity and Real-Time Analysis Experiment: 256-element array status and overview}}.
\newblock \emph{\bibinfo{journal}{Journal of Astronomical Telescopes, Instruments, and Systems}} \textbf{\bibinfo{volume}{8}}, \bibinfo{pages}{011019} (\bibinfo{year}{2022}).

\bibitem{rfof}
\bibinfo{author}{Mena, J.} \emph{et~al.}
\newblock \bibinfo{title}{A radio-frequency-over-fiber link for large-array radio astronomy applications}.
\newblock \emph{\bibinfo{journal}{Journal of Instrumentation}} \textbf{\bibinfo{volume}{8}}, \bibinfo{pages}{T10003–T10003} (\bibinfo{year}{2013}).
\newblock \urlprefix\url{http://dx.doi.org/10.1088/1748-0221/8/10/T10003}.

\bibitem{tm-4}
\bibinfo{author}{Spectrum~Instruments, I.}
\newblock \emph{\bibinfo{title}{Intelligent Reference/TM-4}} (\bibinfo{year}{2008}).
\newblock \urlprefix\url{https://www.spectruminstruments.net/products/tm4/tm4.html}.

\bibitem{2021RNAAS...5..216C}
\bibinfo{author}{{Cary}, S.} \emph{et~al.}
\newblock \bibinfo{title}{{Evaluating and Enhancing Candidate Clocking Systems for CHIME/FRB VLBI Outriggers}}.
\newblock \emph{\bibinfo{journal}{Research Notes of the American Astronomical Society}} \textbf{\bibinfo{volume}{5}}, \bibinfo{pages}{216} (\bibinfo{year}{2021}).

\bibitem{Sanghavi2022frb}
\bibinfo{author}{Sanghavi, P.~R.}
\newblock \emph{\bibinfo{title}{Pathfinding Fast Radio Bursts Localizations Using Very Long Baseline Interferometry}}.
\newblock Ph.D. thesis, \bibinfo{school}{West Virginia University} (\bibinfo{year}{2022}).
\newblock \urlprefix\url{https://www.proquest.com/dissertations-theses/pathfinding-fast-radio-bursts-localizations-using/docview/2734699227/se-2}.
\newblock \bibinfo{note}{Copyright - Database copyright ProQuest LLC; ProQuest does not claim copyright in the individual underlying works; Last updated - 2023-03-06}.

\bibitem{cassanelli2022frb}
\bibinfo{author}{Cassanelli, T.}
\newblock \emph{\bibinfo{title}{Fast Radio Burst Localization with Very Long Baseline Interferometry}}.
\newblock Ph.D. thesis, \bibinfo{school}{University of Toronto} (\bibinfo{year}{2022}).
\newblock \urlprefix\url{https://tspace.library.utoronto.ca/handle/1807/125671}.

\bibitem{leung2023frb}
\bibinfo{author}{Leung, C.}
\newblock \emph{\bibinfo{title}{Localization and Lensing of Fast Radio Bursts using CHIME/FRB and its VLBI Outriggers}}.
\newblock Ph.D. thesis, \bibinfo{school}{Massachusetts Institute of Technology} (\bibinfo{year}{2023}).
\newblock \urlprefix\url{https://hdl.handle.net/1721.1/152953}.

\bibitem{2015arXiv150306189R}
\bibinfo{author}{{Recnik}, A.} \emph{et~al.}
\newblock \bibinfo{title}{{An Efficient Real-time Data Pipeline for the CHIME Pathfinder Radio Telescope X-Engine}}.
\newblock \emph{\bibinfo{journal}{arXiv e-prints}} \bibinfo{pages}{arXiv:1503.06189} (\bibinfo{year}{2015}).

\bibitem{2014SPIE.9145E..22B}
\bibinfo{author}{{Bandura}, K.} \emph{et~al.}
\newblock \bibinfo{editor}{{Stepp}, L.~M.}, \bibinfo{editor}{{Gilmozzi}, R.} \& \bibinfo{editor}{{Hall}, H.~J.} (eds) \emph{\bibinfo{title}{{Canadian Hydrogen Intensity Mapping Experiment (CHIME) pathfinder}}}.
\newblock (eds \bibinfo{editor}{{Stepp}, L.~M.}, \bibinfo{editor}{{Gilmozzi}, R.} \& \bibinfo{editor}{{Hall}, H.~J.}) \emph{\bibinfo{booktitle}{Ground-based and Airborne Telescopes V}}, Vol. \bibinfo{volume}{9145} of \emph{\bibinfo{series}{Society of Photo-Optical Instrumentation Engineers (SPIE) Conference Series}}, \bibinfo{pages}{914522} (\bibinfo{year}{2014}).
\newblock \eprint{1406.2288}.

\bibitem{2014SPIE.9145E..4VN}
\bibinfo{author}{{Newburgh}, L.~B.} \emph{et~al.}
\newblock \bibinfo{editor}{{Stepp}, L.~M.}, \bibinfo{editor}{{Gilmozzi}, R.} \& \bibinfo{editor}{{Hall}, H.~J.} (eds) \emph{\bibinfo{title}{{Calibrating CHIME: a new radio interferometer to probe dark energy}}}.
\newblock (eds \bibinfo{editor}{{Stepp}, L.~M.}, \bibinfo{editor}{{Gilmozzi}, R.} \& \bibinfo{editor}{{Hall}, H.~J.}) \emph{\bibinfo{booktitle}{Ground-based and Airborne Telescopes V}}, Vol. \bibinfo{volume}{9145} of \emph{\bibinfo{series}{Society of Photo-Optical Instrumentation Engineers (SPIE) Conference Series}}, \bibinfo{pages}{91454V} (\bibinfo{year}{2014}).
\newblock \eprint{1406.2267}.

\bibitem{2016ivs..conf..187G}
\bibinfo{author}{{Gordon}, D.}, \bibinfo{author}{{Brisken}, W.} \& \bibinfo{author}{{Max-Moerbeck}, W.}
\newblock \bibinfo{editor}{{Behrend}, D.}, \bibinfo{editor}{{Baver}, K.~D.} \& \bibinfo{editor}{{Armstrong}, K.~L.} (eds) \emph{\bibinfo{title}{{Difxcalc - Calc11 for the DiFX Correlator}}}.
\newblock (eds \bibinfo{editor}{{Behrend}, D.}, \bibinfo{editor}{{Baver}, K.~D.} \& \bibinfo{editor}{{Armstrong}, K.~L.}) \emph{\bibinfo{booktitle}{New Horizons with VGOS}}, \bibinfo{pages}{187--192} (\bibinfo{year}{2016}).

\bibitem{crabvlbi}
\bibinfo{author}{{Lobanov}, A.~P.}, \bibinfo{author}{{Horns}, D.} \& \bibinfo{author}{{Muxlow}, T.~W.~B.}
\newblock \bibinfo{title}{{VLBI imaging of a flare in the Crab nebula: more than just a spot}}.
\newblock \emph{\bibinfo{journal}{\aap}} \textbf{\bibinfo{volume}{533}}, \bibinfo{pages}{A10} (\bibinfo{year}{2011}).

\bibitem{1975MComP..14...55H}
\bibinfo{author}{{Hankins}, T.~H.} \& \bibinfo{author}{{Rickett}, B.~J.}
\newblock \bibinfo{title}{{Pulsar signal processing.}}
\newblock \emph{\bibinfo{journal}{Methods in Computational Physics}} \textbf{\bibinfo{volume}{14}}, \bibinfo{pages}{55--129} (\bibinfo{year}{1975}).

\bibitem{2012hpa..book.....L}
\bibinfo{author}{{Lorimer}, D.~R.} \& \bibinfo{author}{{Kramer}, M.}
\newblock \emph{\bibinfo{title}{{Handbook of Pulsar Astronomy}}}  (\bibinfo{publisher}{Cambridge University Press}, \bibinfo{year}{2012}).

\bibitem{2020arXiv200702886K}
\bibinfo{author}{{Kulkarni}, S.~R.}
\newblock \bibinfo{title}{{Dispersion measure: Confusion, Constants \& Clarity}}.
\newblock \emph{\bibinfo{journal}{arXiv e-prints}} \bibinfo{pages}{arXiv:2007.02886} (\bibinfo{year}{2020}).

\bibitem{2022ApJ...927L...3N}
\bibinfo{author}{{Nimmo}, K.} \emph{et~al.}
\newblock \bibinfo{title}{{Milliarcsecond Localization of the Repeating FRB 20201124A}}.
\newblock \emph{\bibinfo{journal}{\apjl}} \textbf{\bibinfo{volume}{927}}, \bibinfo{pages}{L3} (\bibinfo{year}{2022}).

\bibitem{1998AJ....116..516M}
\bibinfo{author}{{Ma}, C.} \emph{et~al.}
\newblock \bibinfo{title}{{The International Celestial Reference Frame as Realized by Very Long Baseline Interferometry}}.
\newblock \emph{\bibinfo{journal}{\aj}} \textbf{\bibinfo{volume}{116}}, \bibinfo{pages}{516--546} (\bibinfo{year}{1998}).

\bibitem{2021ARA&A..59...59B}
\bibinfo{author}{{Brown}, A. G.~A.}
\newblock \bibinfo{title}{{Microarcsecond Astrometry: Science Highlights from Gaia}}.
\newblock \emph{\bibinfo{journal}{\araa}} \textbf{\bibinfo{volume}{59}}, \bibinfo{pages}{59--115} (\bibinfo{year}{2021}).

\bibitem{2005USNOC.179.....K}
\bibinfo{author}{{Kaplan}, G.~H.}
\newblock \bibinfo{title}{{The IAU resolutions on astronomical reference systems, time scales, and earth rotation models : explanation and implementation}}.
\newblock \emph{\bibinfo{journal}{U.S. Naval Observatory Circulars}} \textbf{\bibinfo{volume}{179}} (\bibinfo{year}{2005}).

\bibitem{1998A&A...331L..33F}
\bibinfo{author}{{Feissel}, M.} \& \bibinfo{author}{{Mignard}, F.}
\newblock \bibinfo{title}{{The adoption of ICRS on 1 January 1998: meaning and consequences}}.
\newblock \emph{\bibinfo{journal}{\aap}} \textbf{\bibinfo{volume}{331}}, \bibinfo{pages}{L33--L36} (\bibinfo{year}{1998}).

\bibitem{2019ascl.soft10004S}
\bibinfo{author}{{Seymour}, A.}, \bibinfo{author}{{Michilli}, D.} \& \bibinfo{author}{{Pleunis}, Z.}
\newblock \bibinfo{title}{{DM\_phase: Algorithm for correcting dispersion of radio signals}}.
\newblock \bibinfo{howpublished}{Astrophysics Source Code Library, record ascl:1910.004} (\bibinfo{year}{2019}).
\newblock \eprint{1910.004}.

\bibitem{2024ApJS..271...49F}
\bibinfo{author}{{Fonseca}, E.} \emph{et~al.}
\newblock \bibinfo{title}{{Modeling the Morphology of Fast Radio Bursts and Radio Pulsars with fitburst}}.
\newblock \emph{\bibinfo{journal}{\apjs}} \textbf{\bibinfo{volume}{271}}, \bibinfo{pages}{49} (\bibinfo{year}{2024}).

\bibitem{ymw16}
\bibinfo{author}{{Yao}, J.~M.}, \bibinfo{author}{{Manchester}, R.~N.} \& \bibinfo{author}{{Wang}, N.}
\newblock \bibinfo{title}{{A New Electron-density Model for Estimation of Pulsar and FRB Distances}}.
\newblock \emph{\bibinfo{journal}{\apj}} \textbf{\bibinfo{volume}{835}}, \bibinfo{pages}{29} (\bibinfo{year}{2017}).

\bibitem{2020ApJ...888..105Y}
\bibinfo{author}{{Yamasaki}, S.} \& \bibinfo{author}{{Totani}, T.}
\newblock \bibinfo{title}{{The Galactic Halo Contribution to the Dispersion Measure of Extragalactic Fast Radio Bursts}}.
\newblock \emph{\bibinfo{journal}{\apj}} \textbf{\bibinfo{volume}{888}}, \bibinfo{pages}{105} (\bibinfo{year}{2020}).

\bibitem{2020MNRAS.496L.106K}
\bibinfo{author}{{Keating}, L.~C.} \& \bibinfo{author}{{Pen}, U.-L.}
\newblock \bibinfo{title}{{Exploring the dispersion measure of the Milky Way halo}}.
\newblock \emph{\bibinfo{journal}{\mnras}} \textbf{\bibinfo{volume}{496}}, \bibinfo{pages}{L106--L110} (\bibinfo{year}{2020}).

\bibitem{2023ApJ...946...58C}
\bibinfo{author}{{Cook}, A.~M.} \emph{et~al.}
\newblock \bibinfo{title}{{An FRB Sent Me a DM: Constraining the Electron Column of the Milky Way Halo with Fast Radio Burst Dispersion Measures from CHIME/FRB}}.
\newblock \emph{\bibinfo{journal}{\apj}} \textbf{\bibinfo{volume}{946}}, \bibinfo{pages}{58} (\bibinfo{year}{2023}).

\bibitem{2020Natur.581..391M}
\bibinfo{author}{{Macquart}, J.~P.} \emph{et~al.}
\newblock \bibinfo{title}{{A census of baryons in the Universe from localized fast radio bursts}}.
\newblock \emph{\bibinfo{journal}{\nat}} \textbf{\bibinfo{volume}{581}}, \bibinfo{pages}{391--395} (\bibinfo{year}{2020}).

\bibitem{2021MNRAS.505.5356B}
\bibinfo{author}{{Batten}, A.~J.} \emph{et~al.}
\newblock \bibinfo{title}{{The cosmic dispersion measure in the EAGLE simulations}}.
\newblock \emph{\bibinfo{journal}{\mnras}} \textbf{\bibinfo{volume}{505}}, \bibinfo{pages}{5356--5369} (\bibinfo{year}{2021}).

\bibitem{2020MNRAS.493...87T}
\bibinfo{author}{{Trujillo}, I.}, \bibinfo{author}{{Chamba}, N.} \& \bibinfo{author}{{Knapen}, J.~H.}
\newblock \bibinfo{title}{{A physically motivated definition for the size of galaxies in an era of ultradeep imaging}}.
\newblock \emph{\bibinfo{journal}{\mnras}} \textbf{\bibinfo{volume}{493}}, \bibinfo{pages}{87--105} (\bibinfo{year}{2020}).

\bibitem{2015ApJ...806...96L}
\bibinfo{author}{{Licquia}, T.~C.} \& \bibinfo{author}{{Newman}, J.~A.}
\newblock \bibinfo{title}{{Improved Estimates of the Milky Way's Stellar Mass and Star Formation Rate from Hierarchical Bayesian Meta-Analysis}}.
\newblock \emph{\bibinfo{journal}{\apj}} \textbf{\bibinfo{volume}{806}}, \bibinfo{pages}{96} (\bibinfo{year}{2015}).

\bibitem{2020ApJ...897..124O}
\bibinfo{author}{{Ocker}, S.~K.}, \bibinfo{author}{{Cordes}, J.~M.} \& \bibinfo{author}{{Chatterjee}, S.}
\newblock \bibinfo{title}{{Electron Density Structure of the Local Galactic Disk}}.
\newblock \emph{\bibinfo{journal}{\apj}} \textbf{\bibinfo{volume}{897}}, \bibinfo{pages}{124} (\bibinfo{year}{2020}).

\bibitem{2021ApJ...911..102O}
\bibinfo{author}{{Ocker}, S.~K.}, \bibinfo{author}{{Cordes}, J.~M.} \& \bibinfo{author}{{Chatterjee}, S.}
\newblock \bibinfo{title}{{Constraining Galaxy Halos from the Dispersion and Scattering of Fast Radio Bursts and Pulsars}}.
\newblock \emph{\bibinfo{journal}{\apj}} \textbf{\bibinfo{volume}{911}}, \bibinfo{pages}{102} (\bibinfo{year}{2021}).

\bibitem{2016arXiv160505890C}
\bibinfo{author}{{Cordes}, J.~M.}, \bibinfo{author}{{Wharton}, R.~S.}, \bibinfo{author}{{Spitler}, L.~G.}, \bibinfo{author}{{Chatterjee}, S.} \& \bibinfo{author}{{Wasserman}, I.}
\newblock \bibinfo{title}{{Radio Wave Propagation and the Provenance of Fast Radio Bursts}}.
\newblock \emph{\bibinfo{journal}{arXiv e-prints}} \bibinfo{pages}{arXiv:1605.05890} (\bibinfo{year}{2016}).

\bibitem{2022MNRAS.509.4775J}
\bibinfo{author}{{James}, C.~W.} \emph{et~al.}
\newblock \bibinfo{title}{{The z-DM distribution of fast radio bursts}}.
\newblock \emph{\bibinfo{journal}{\mnras}} \textbf{\bibinfo{volume}{509}}, \bibinfo{pages}{4775--4802} (\bibinfo{year}{2022}).

\bibitem{2014ApJS..212....6O}
\bibinfo{author}{{Olausen}, S.~A.} \& \bibinfo{author}{{Kaspi}, V.~M.}
\newblock \bibinfo{title}{{The McGill Magnetar Catalog}}.
\newblock \emph{\bibinfo{journal}{\apjs}} \textbf{\bibinfo{volume}{212}}, \bibinfo{pages}{6} (\bibinfo{year}{2014}).

\bibitem{1979ApJS...41..513M}
\bibinfo{author}{{Miller}, G.~E.} \& \bibinfo{author}{{Scalo}, J.~M.}
\newblock \bibinfo{title}{{The Initial Mass Function and Stellar Birthrate in the Solar Neighborhood}}.
\newblock \emph{\bibinfo{journal}{\apjs}} \textbf{\bibinfo{volume}{41}}, \bibinfo{pages}{513} (\bibinfo{year}{1979}).

\bibitem{fab+20}
\bibinfo{author}{{Fonseca}, E.} \emph{et~al.}
\newblock \bibinfo{title}{{Nine New Repeating Fast Radio Burst Sources from CHIME/FRB}}.
\newblock \emph{\bibinfo{journal}{\apjl}} \textbf{\bibinfo{volume}{891}}, \bibinfo{pages}{L6} (\bibinfo{year}{2020}).

\bibitem{2021ApJ...910L..18B}
\bibinfo{author}{{Bhardwaj}, M.} \emph{et~al.}
\newblock \bibinfo{title}{{A Nearby Repeating Fast Radio Burst in the Direction of M81}}.
\newblock \emph{\bibinfo{journal}{\apjl}} \textbf{\bibinfo{volume}{910}}, \bibinfo{pages}{L18} (\bibinfo{year}{2021}).

\bibitem{b66}
\bibinfo{author}{{Burn}, B.~J.}
\newblock \bibinfo{title}{{On the depolarization of discrete radio sources by Faraday dispersion}}.
\newblock \emph{\bibinfo{journal}{\mnras}} \textbf{\bibinfo{volume}{133}}, \bibinfo{pages}{67} (\bibinfo{year}{1966}).

\bibitem{bb05}
\bibinfo{author}{{Brentjens}, M.~A.} \& \bibinfo{author}{{de Bruyn}, A.~G.}
\newblock \bibinfo{title}{{Faraday rotation measure synthesis}}.
\newblock \emph{\bibinfo{journal}{\aap}} \textbf{\bibinfo{volume}{441}}, \bibinfo{pages}{1217--1228} (\bibinfo{year}{2005}).

\bibitem{Mckinven_2021}
\bibinfo{author}{Mckinven, R.} \emph{et~al.}
\newblock \bibinfo{title}{Polarization pipeline for fast radio bursts detected by {CHIME}/{FRB}}.
\newblock \emph{\bibinfo{journal}{The Astrophysical Journal}} \textbf{\bibinfo{volume}{920}}, \bibinfo{pages}{138} (\bibinfo{year}{2021}).
\newblock \urlprefix\url{https://doi.org/10.3847/1538-4357/ac126a}.

\bibitem{Vedantham2019}
\bibinfo{author}{{Vedantham}, H.~K.} \& \bibinfo{author}{{Ravi}, V.}
\newblock \bibinfo{title}{{Faraday conversion and magneto-ionic variations in fast radio bursts}}.
\newblock \emph{\bibinfo{journal}{\mnras}} \textbf{\bibinfo{volume}{485}}, \bibinfo{pages}{L78--L82} (\bibinfo{year}{2019}).

\bibitem{Gruzinov2019}
\bibinfo{author}{{Gruzinov}, A.} \& \bibinfo{author}{{Levin}, Y.}
\newblock \bibinfo{title}{{Conversion Measure of Faraday Rotation-Conversion with Application to Fast Radio Bursts}}.
\newblock \emph{\bibinfo{journal}{\apj}} \textbf{\bibinfo{volume}{876}}, \bibinfo{pages}{74} (\bibinfo{year}{2019}).

\bibitem{Beniamini2022}
\bibinfo{author}{{Beniamini}, P.}, \bibinfo{author}{{Kumar}, P.} \& \bibinfo{author}{{Narayan}, R.}
\newblock \bibinfo{title}{{Faraday depolarization and induced circular polarization by multipath propagation with application to FRBs}}.
\newblock \emph{\bibinfo{journal}{\mnras}} \textbf{\bibinfo{volume}{510}}, \bibinfo{pages}{4654--4668} (\bibinfo{year}{2022}).

\bibitem{Hutschenreuter2021}
\bibinfo{author}{{Hutschenreuter}, S.} \emph{et~al.}
\newblock \bibinfo{title}{{The Galactic Faraday rotation sky 2020}}.
\newblock \emph{\bibinfo{journal}{\aap}} \textbf{\bibinfo{volume}{657}}, \bibinfo{pages}{A43} (\bibinfo{year}{2022}).

\bibitem{Mevius2018b}
\bibinfo{author}{{Mevius}, M.}
\newblock \bibinfo{title}{{RMextract: Ionospheric Faraday Rotation calculator}} (\bibinfo{year}{2018}).
\newblock \eprint{1806.024}.

\bibitem{Manchester2005}
\bibinfo{author}{{Manchester}, R.~N.}, \bibinfo{author}{{Hobbs}, G.~B.}, \bibinfo{author}{{Teoh}, A.} \& \bibinfo{author}{{Hobbs}, M.}
\newblock \bibinfo{title}{{The Australia Telescope National Facility Pulsar Catalogue}}.
\newblock \emph{\bibinfo{journal}{\aj}} \textbf{\bibinfo{volume}{129}}, \bibinfo{pages}{1993--2006} (\bibinfo{year}{2005}).

\bibitem{psrqpy}
\bibinfo{author}{{Pitkin}, M.}
\newblock \bibinfo{title}{{psrqpy: a python interface for querying the ATNF pulsar catalogue}}.
\newblock \emph{\bibinfo{journal}{{Journal of Open Source Software}}} \textbf{\bibinfo{volume}{3}}, \bibinfo{pages}{538} (\bibinfo{year}{2018}).
\newblock \urlprefix\url{https://doi.org/10.21105/joss.00538}.

\bibitem{2004Magnier}
\bibinfo{author}{{Magnier}, E.~A.} \& \bibinfo{author}{{Cuillandre}, J.~C.}
\newblock \bibinfo{title}{{The Elixir System: Data Characterization and Calibration at the Canada-France-Hawaii Telescope}}.
\newblock \emph{\bibinfo{journal}{\pasp}} \textbf{\bibinfo{volume}{116}}, \bibinfo{pages}{449--464} (\bibinfo{year}{2004}).

\bibitem{2013Prunet}
\bibinfo{author}{{Prunet}, S.}, \bibinfo{author}{{Fouque}, P.} \& \bibinfo{author}{{Gwyn}, S.}
\newblock \emph{\bibinfo{title}{{Photometric calibration of Megacam data}}} (\bibinfo{year}{2014}).

\bibitem{PetroRadii}
\bibinfo{author}{Graham, A.~W.} \emph{et~al.}
\newblock \bibinfo{title}{Total galaxy magnitudes and effective radii from petrosian magnitudes and radii}.
\newblock \emph{\bibinfo{journal}{The Astronomical Journal}} \textbf{\bibinfo{volume}{130}}, \bibinfo{pages}{1535–1544} (\bibinfo{year}{2005}).
\newblock \urlprefix\url{http://dx.doi.org/10.1086/444475}.

\bibitem{2007FM}
\bibinfo{author}{{Fitzpatrick}, E.~L.} \& \bibinfo{author}{{Massa}, D.}
\newblock \bibinfo{title}{{An Analysis of the Shapes of Interstellar Extinction Curves. V. The IR-through-UV Curve Morphology}}.
\newblock \emph{\bibinfo{journal}{\apj}} \textbf{\bibinfo{volume}{663}}, \bibinfo{pages}{320--341} (\bibinfo{year}{2007}).

\bibitem{2010MNRAS.405.1409C}
\bibinfo{author}{{Chilingarian}, I.~V.}, \bibinfo{author}{{Melchior}, A.-L.} \& \bibinfo{author}{{Zolotukhin}, I.~Y.}
\newblock \bibinfo{title}{{Analytical approximations of K-corrections in optical and near-infrared bands}}.
\newblock \emph{\bibinfo{journal}{\mnras}} \textbf{\bibinfo{volume}{405}}, \bibinfo{pages}{1409--1420} (\bibinfo{year}{2010}).

\bibitem{1986IRAF}
\bibinfo{author}{{Tody}, D.}
\newblock \bibinfo{editor}{{Crawford}, D.~L.} (ed.) \emph{\bibinfo{title}{{The IRAF Data Reduction and Analysis System}}}.
\newblock (ed.\bibinfo{editor}{{Crawford}, D.~L.}) \emph{\bibinfo{booktitle}{Instrumentation in astronomy VI}}, Vol. \bibinfo{volume}{627} of \emph{\bibinfo{series}{Society of Photo-Optical Instrumentation Engineers (SPIE) Conference Series}}, \bibinfo{pages}{733} (\bibinfo{year}{1986}).

\bibitem{1993IRAF}
\bibinfo{author}{{Tody}, D.}
\newblock \bibinfo{editor}{{Hanisch}, R.~J.}, \bibinfo{editor}{{Brissenden}, R.~J.~V.} \& \bibinfo{editor}{{Barnes}, J.} (eds) \emph{\bibinfo{title}{{IRAF in the Nineties}}}.
\newblock (eds \bibinfo{editor}{{Hanisch}, R.~J.}, \bibinfo{editor}{{Brissenden}, R.~J.~V.} \& \bibinfo{editor}{{Barnes}, J.}) \emph{\bibinfo{booktitle}{Astronomical Data Analysis Software and Systems II}}, Vol.~\bibinfo{volume}{52} of \emph{\bibinfo{series}{Astronomical Society of the Pacific Conference Series}}, \bibinfo{pages}{173} (\bibinfo{year}{1993}).

\bibitem{Bhardwaj_2021}
\bibinfo{author}{Bhardwaj, M.} \emph{et~al.}
\newblock \bibinfo{title}{A local universe host for the repeating fast radio burst {FRB} 20181030a}.
\newblock \emph{\bibinfo{journal}{The Astrophysical Journal Letters}} \textbf{\bibinfo{volume}{919}}, \bibinfo{pages}{L24} (\bibinfo{year}{2021}).
\newblock \urlprefix\url{https://doi.org/10.3847/2041-8213/ac223b}.

\bibitem{Shao_2007}
\bibinfo{author}{Shao, Z.} \emph{et~al.}
\newblock \bibinfo{title}{Inclination-dependent luminosity function of spiral galaxies in the sloan digital sky survey: Implications for dust extinction}.
\newblock \emph{\bibinfo{journal}{The Astrophysical Journal}} \textbf{\bibinfo{volume}{659}}, \bibinfo{pages}{1159--1171} (\bibinfo{year}{2007}).
\newblock \urlprefix\url{https://doi.org/10.1086/511131}.

\bibitem{2000calzetti}
\bibinfo{author}{{Calzetti}, D.} \emph{et~al.}
\newblock \bibinfo{title}{{The Dust Content and Opacity of Actively Star-forming Galaxies}}.
\newblock \emph{\bibinfo{journal}{\apj}} \textbf{\bibinfo{volume}{533}}, \bibinfo{pages}{682--695} (\bibinfo{year}{2000}).

\bibitem{2017Byler}
\bibinfo{author}{{Byler}, N.}, \bibinfo{author}{{Dalcanton}, J.~J.}, \bibinfo{author}{{Conroy}, C.} \& \bibinfo{author}{{Johnson}, B.~D.}
\newblock \bibinfo{title}{{Nebular Continuum and Line Emission in Stellar Population Synthesis Models}}.
\newblock \emph{\bibinfo{journal}{\apj}} \textbf{\bibinfo{volume}{840}}, \bibinfo{pages}{44} (\bibinfo{year}{2017}).

\bibitem{2007DraineLi}
\bibinfo{author}{{Draine}, B.~T.} \& \bibinfo{author}{{Li}, A.}
\newblock \bibinfo{title}{{Infrared Emission from Interstellar Dust. IV. The Silicate-Graphite-PAH Model in the Post-Spitzer Era}}.
\newblock \emph{\bibinfo{journal}{\apj}} \textbf{\bibinfo{volume}{657}}, \bibinfo{pages}{810--837} (\bibinfo{year}{2007}).

\bibitem{2010Bernardi}
\bibinfo{author}{{Bernardi}, M.} \emph{et~al.}
\newblock \bibinfo{title}{{Galaxy luminosities, stellar masses, sizes, velocity dispersions as a function of morphological type}}.
\newblock \emph{\bibinfo{journal}{\mnras}} \textbf{\bibinfo{volume}{404}}, \bibinfo{pages}{2087--2122} (\bibinfo{year}{2010}).

\bibitem{1994ApJ...435...22K}
\bibinfo{author}{{Kennicutt}, J., Robert~C.}, \bibinfo{author}{{Tamblyn}, P.} \& \bibinfo{author}{{Congdon}, C.~E.}
\newblock \bibinfo{title}{{Past and Future Star Formation in Disk Galaxies}}.
\newblock \emph{\bibinfo{journal}{\apj}} \textbf{\bibinfo{volume}{435}}, \bibinfo{pages}{22} (\bibinfo{year}{1994}).

\bibitem{1999Fitz}
\bibinfo{author}{{Fitzpatrick}, E.~L.}
\newblock \bibinfo{title}{{Correcting for the Effects of Interstellar Extinction}}.
\newblock \emph{\bibinfo{journal}{\pasp}} \textbf{\bibinfo{volume}{111}}, \bibinfo{pages}{63--75} (\bibinfo{year}{1999}).

\bibitem{conroy2009a}
\bibinfo{author}{{Conroy}, C.}, \bibinfo{author}{{Gunn}, J.~E.} \& \bibinfo{author}{{White}, M.}
\newblock \bibinfo{title}{{The Propagation of Uncertainties in Stellar Population Synthesis Modeling. I. The Relevance of Uncertain Aspects of Stellar Evolution and the Initial Mass Function to the Derived Physical Properties of Galaxies}}.
\newblock \emph{\bibinfo{journal}{\apj}} \textbf{\bibinfo{volume}{699}}, \bibinfo{pages}{486--506} (\bibinfo{year}{2009}).

\bibitem{2011FS}
\bibinfo{author}{{Schlafly}, E.~F.} \& \bibinfo{author}{{Finkbeiner}, D.~P.}
\newblock \bibinfo{title}{{Measuring Reddening with Sloan Digital Sky Survey Stellar Spectra and Recalibrating SFD}}.
\newblock \emph{\bibinfo{journal}{\apj}} \textbf{\bibinfo{volume}{737}}, \bibinfo{pages}{103} (\bibinfo{year}{2011}).

\bibitem{o_donnell1994r}
\bibinfo{author}{{O'Donnell}, J.~E.}
\newblock \bibinfo{title}{{R v-dependent Optical and Near-Ultraviolet Extinction}}.
\newblock \emph{\bibinfo{journal}{\apj}} \textbf{\bibinfo{volume}{422}}, \bibinfo{pages}{158} (\bibinfo{year}{1994}).

\bibitem{2022Natur.602..585K}
\bibinfo{author}{{Kirsten}, F.} \emph{et~al.}
\newblock \bibinfo{title}{{A repeating fast radio burst source in a globular cluster}}.
\newblock \emph{\bibinfo{journal}{\nat}} \textbf{\bibinfo{volume}{602}}, \bibinfo{pages}{585--589} (\bibinfo{year}{2022}).

\bibitem{2006MNRAS.369.1688D}
\bibinfo{author}{{Dehnen}, W.}, \bibinfo{author}{{McLaughlin}, D.~E.} \& \bibinfo{author}{{Sachania}, J.}
\newblock \bibinfo{title}{{The velocity dispersion and mass profile of the Milky Way}}.
\newblock \emph{\bibinfo{journal}{\mnras}} \textbf{\bibinfo{volume}{369}}, \bibinfo{pages}{1688--1692} (\bibinfo{year}{2006}).

\bibitem{2018AJ....156..123A}
\bibinfo{author}{{Astropy Collaboration}} \emph{et~al.}
\newblock \bibinfo{title}{{The Astropy Project: Building an Open-science Project and Status of the v2.0 Core Package}}.
\newblock \emph{\bibinfo{journal}{\aj}} \textbf{\bibinfo{volume}{156}}, \bibinfo{pages}{123} (\bibinfo{year}{2018}).

\bibitem{marten_van_kerkwijk_2020_4292543}
\bibinfo{author}{van Kerkwijk, M.} \emph{et~al.}
\newblock \bibinfo{title}{mhvk/baseband: v4.1.3} (\bibinfo{year}{2023}).
\newblock \urlprefix\url{https://doi.org/10.5281/zenodo.7941170}.

\bibitem{2007CSE.....9...90H}
\bibinfo{author}{{Hunter}, J.~D.}
\newblock \bibinfo{title}{{Matplotlib: A 2D Graphics Environment}}.
\newblock \emph{\bibinfo{journal}{Computing in Science and Engineering}} \textbf{\bibinfo{volume}{9}}, \bibinfo{pages}{90--95} (\bibinfo{year}{2007}).

\bibitem{2020Natur.585..357H}
\bibinfo{author}{{Harris}, C.~R.} \emph{et~al.}
\newblock \bibinfo{title}{{Array programming with NumPy}}.
\newblock \emph{\bibinfo{journal}{\nat}} \textbf{\bibinfo{volume}{585}}, \bibinfo{pages}{357--362} (\bibinfo{year}{2020}).

\bibitem{2020SciPy-NMeth}
\bibinfo{author}{Virtanen, P.} \emph{et~al.}
\newblock \bibinfo{title}{{{SciPy} 1.0: Fundamental Algorithms for Scientific Computing in Python}}.
\newblock \emph{\bibinfo{journal}{Nature Methods}} \textbf{\bibinfo{volume}{17}}, \bibinfo{pages}{261--272} (\bibinfo{year}{2020}).

\bibitem{hdf5}
\bibinfo{author}{Collette, A.} \emph{et~al.}
\newblock \bibinfo{title}{h5py/h5py: 3.2.1} (\bibinfo{year}{2021}).
\newblock \urlprefix\url{https://zenodo.org/record/4584676}.

\bibitem{Foreman-Mackey2016}
\bibinfo{author}{Foreman-Mackey, D.}
\newblock \bibinfo{title}{corner.py: Scatterplot matrices in python}.
\newblock \emph{\bibinfo{journal}{Journal of Open Source Software}} \textbf{\bibinfo{volume}{1}}, \bibinfo{pages}{24} (\bibinfo{year}{2016}).
\newblock \urlprefix\url{https://doi.org/10.21105/joss.00024}.

\bibitem{Cartopy}
\bibinfo{author}{{Met Office}}.
\newblock \emph{\bibinfo{title}{Cartopy: a cartographic python library with a matplotlib interface}}.
\newblock \bibinfo{address}{Exeter, Devon} (\bibinfo{year}{2015}).
\newblock \urlprefix\url{http://scitools.org.uk/cartopy}.

\end{thebibliography}

\end{document}